\documentclass[pra,superscriptaddress,a4paper,twocolumn,10pt]{revtex4-2}
\usepackage{physics}
\usepackage{amsmath,amssymb}
\usepackage{braket}
\usepackage{graphicx}
\usepackage{xcolor}
\usepackage{verbatim}
\usepackage{color}
\usepackage[colorlinks, linkcolor={blue}, citecolor={blue}, urlcolor={blue}]{hyperref}
\usepackage{tabularx}
\usepackage{booktabs}
\usepackage{blindtext}
\usepackage{epstopdf}
\usepackage{nicefrac}
\usepackage{mathrsfs}
\usepackage{orcidlink}

\renewcommand{\vec}{\mathbf}

\begin{document}
\title{Growing Extended Laughlin States in a Quantum Gas Microscope: \\ A Patchwork Construction}
\author{F.~A.~Palm\orcidlink{0000-0001-5774-5546}}
\affiliation{Department of Physics and Arnold Sommerfeld Center for Theoretical Physics (ASC), Ludwig-Maximilians-Universit\"at M\"unchen, Theresienstr. 37, D-80333 M\"unchen, Germany}
\affiliation{Munich Center for Quantum Science and Technology (MCQST), Schellingstr. 4, D-80799 M\"unchen, Germany}
\affiliation{CENOLI, Universit\'e Libre de Bruxelles, CP 231, Campus Plaine, B-1050 Brussels, Belgium}
\author{J.~Kwan}
\affiliation{Department of Physics, Harvard University, Cambridge, Massachusetts 02138, USA}
\author{B.~Bakkali-Hassani}
\affiliation{Department of Physics, Harvard University, Cambridge, Massachusetts 02138, USA}
\author{M.~Greiner\orcidlink{0000-0002-2935-2363}}
\affiliation{Department of Physics, Harvard University, Cambridge, Massachusetts 02138, USA}
\author{U.~Schollw\"ock\orcidlink{0000-0002-2538-1802}}
\affiliation{Department of Physics and Arnold Sommerfeld Center for Theoretical Physics (ASC), Ludwig-Maximilians-Universit\"at M\"unchen, Theresienstr. 37, D-80333 M\"unchen, Germany}
\affiliation{Munich Center for Quantum Science and Technology (MCQST), Schellingstr. 4, D-80799 M\"unchen, Germany}
\author{N.~Goldman}
\affiliation{CENOLI, Universit\'e Libre de Bruxelles, CP 231, Campus Plaine, B-1050 Brussels, Belgium}
\affiliation{Laboratoire Kastler Brossel, Coll\`ege de France, CNRS, ENS-Universit\'e PSL, Sorbonne Universit\'e, 11 Place Marcelin Berthelot, 75005 Paris, France}
\author{F.~Grusdt\orcidlink{0000-0003-3531-8089}}
\affiliation{Department of Physics and Arnold Sommerfeld Center for Theoretical Physics (ASC), Ludwig-Maximilians-Universit\"at M\"unchen, Theresienstr. 37, D-80333 M\"unchen, Germany}
\affiliation{Munich Center for Quantum Science and Technology (MCQST), Schellingstr. 4, D-80799 M\"unchen, Germany}
\date{\today}
\begin{abstract}
	The study of fractional Chern insulators and their exotic anyonic excitations poses a major challenge in current experimental and theoretical research.
	Quantum simulators, in particular ultracold atoms in optical lattices, provide a promising platform to realize, manipulate, and understand such systems with a high degree of controllability.
	Recently, an atomic $ \nu=\nicefrac{1}{2}$ Laughlin state has been realized experimentally for a small system of two particles on $4\times4$ sites [L\'eonard et al., Nature (2023)].
	The next challenge concerns the preparation of Laughlin states in extended systems, ultimately giving access to anyonic braiding statistics or gapless chiral edge-states in systems with open boundaries.
	Here, we propose and analyze an experimentally feasible scheme to grow larger Laughlin states by connecting multiple copies of the already existing $4\times4$-system.
	First, we present a minimal setting obtained by coupling two of such patches, producing an extended $8\times4$-system with four particles.
	Then, we analyze different preparation schemes, setting the focus on two shapes for the extended system, and discuss their respective advantages:
	While growing strip-like lattices could give experimental access to the central charge, square-like geometries are advantageous for creating quasi-hole excitations in view of braiding protocols.
	We highlight the robust quantization of the fractional quasi-hole charge upon using our preparation protocol.
	We benchmark the performance of our patchwork preparation scheme by comparing it to a protocol based on coupling one-dimensional chains.
	We find that the patchwork approach consistently gives higher target-state fidelities, especially for elongated systems.
	The results presented here pave the way towards near-term implementations of extended Laughlin states in quantum gas microscopes and the subsequent exploration of exotic properties of topologically ordered systems in experiments.
\end{abstract}
\maketitle
\section{Introduction}

\begin{figure*}[!t]
	\centering
	\includegraphics{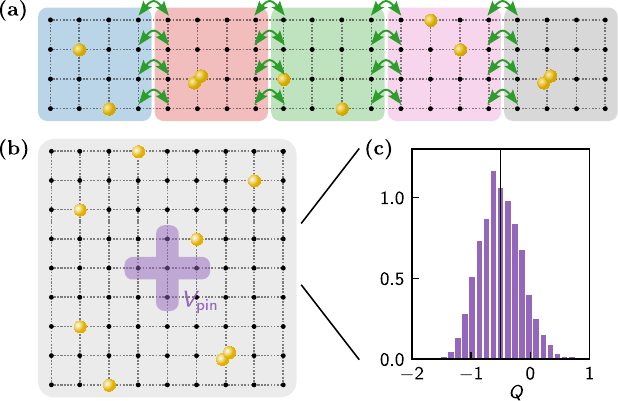}
	\caption{
		\textbf{(a)} Schematic of the proposed patchwork scheme to grow extended Laughlin states for the case of five patches coupled in a super-chain.
		\textbf{(b)} Visualization of the setup pinning a fractionally charged quasi-hole as obtained by coupling four patches in a large square geometry.
		\textbf{(c)} The full counting statistics of the pinned charge, defined using a smooth envelope function described later in this paper, clearly shows the quantization of the fractional charge to $q_{\rm qh}=-\nicefrac{1}{2}$.
	}
	\label{fig:Fig1}
\end{figure*}

Interacting topological states of matter constitute an exotic class of quantum phases with potential applications in topological quantum computation.
Fractional quantum Hall states~\cite{Tsui1982} are a paradigmatic family of such states, exhibiting properties like quantum number fractionalization, manifesting in fractionally charged quasi-particles with anyonic braiding statistics~\cite{Laughlin1983,Halperin1984,Arovas1984}.
In recent years, quantum simulation platforms utilizing cold atoms in optical lattices subject to artificial magnetic fields have proved useful to realize topological phases in a controlled and tunable setup~\cite{Aidelsburger2013,Miyake2013,Tai2017,Jotzu2014,Flaeschner2016,Goldman2016,Aidelsburger2018,Cooper2019}.

Despite recent advances, quantum simulation of Laughlin states has so far been limited to systems of two photons~\cite{Clark2020} or, very recently, two bosonic atoms forming a fractional Chern insulator (FCI) in an optical lattice~\cite{Leonard2023a}.
Realizing fractional quantum Hall states of many particles is desirable, in view of their further exploration in quantum simulators, however this task remains extremely challenging.
Specifically, in the thermodynamic limit topologically ordered states are separated from trivial ones by a gap closing and therefore adiabatic preparation schemes are hardly available in large systems.
In contrast, small systems exhibit a significant finite size gap, allowing for the adiabatic preparation of topologically ordered states from trivial states~\cite{Popp2004,Barkeshli2014,Barkeshli2015,Motruk2017,Hudomal2019,Andrade2021,He2017}.

Experiments ultimately aim for genuine many-body states in large systems.
Among the main goals of cold atom experiments on FCIs is the direct observation of (non-Abelian) anyon braiding, which necessarily requires enough particles to realize quasi-particle or quasi-hole states as well as large enough systems to move the particles around each other in real space~\cite{Grusdt2016,Nakamura2020}.
Alternative proposals to extract the braiding statistics from density measurements also require large systems compared to those available in existing experiments~\cite{Umucalilar2018,Macaluso2020}, as well as proposals to directly measure the central charge in elongated systems~\cite{Palm2022}.
Recently, it was proposed to perform edge mode spectroscopy on FCIs, where sufficiently large particle numbers and system sizes even allow for the resolution of the magneto-roton mode along with improved resolution of the chiral edge mode~\cite{Binanti2023}.
Therefore, it is of prime importance to develop protocols that take the existing experiments to their full potential by growing large systems.

In general, there exist various approaches to prepare topologically ordered states:
Some proposals suggested to directly cross a continuous phase transition between a trivial and a topological state~\cite{Popp2004,Barkeshli2014,Barkeshli2015,Motruk2017,Hudomal2019,Andrade2021}, or to utilize a coupled-wire paradigm~\cite{Kane2002,Teo2014} and exploit the close similarity between weakly coupled Luttinger liquids and fractional quantum Hall states~\cite{He2017}.
In contrast, step-wise growing schemes - like the one proposed in this work - start from a topological few-particle state to which particles are added step-by-step at the edge~\cite{Grusdt2014a,Homeier2021,Liu2022a}.
Recently, a novel protocol using a reservoir approach has been proposed to grow interacting topological states in optical box potentials~\cite{Wang2023}.
Finally, dissipative state preparation is especially promising in the context of superconducting qubits~\cite{Kapit2014,Liu2021a} and photonic systems~\cite{Kurilovich2022}.

Here, we propose to couple existing Laughlin state patches of two bosons on $4\times 4$ sites~\cite{Leonard2023a} to grow extended FCIs.
We perform proof-of-principle simulations to first connect two patches employing different protocols, varying local hopping amplitudes or potential barriers.
Afterwards, we extend our protocol to grow extended chains of up to five patches hosting ten particles in total, thus ending up with a total system size of $20\times 4$ sites, see Fig.~\ref{fig:Fig1}\textbf{(a)}.
Such elongated systems were found to exhibit gapless chiral edge modes despite their finite size along the short direction~\cite{Palm2022}.
We compare different protocols for coupling the patches and provide evidence for the scalability of our approach.
Finally, we connect four patches in a square geometry of $9 \times 9$ sites hosting eight bosons, allowing for the direct preparation of a large Laughlin state in a geometry which is promising for future braiding experiments.
Furthermore, we create a quasi-hole state by adding a local potential (Fig.~\ref{fig:Fig1}\textbf{(b)}) and for the first time directly confirm the fractionalization of the quasi-hole charge using the full counting statistics of the pinned charge, see Fig.~\ref{fig:Fig1}\textbf{(c)}.
We demonstrate that this state can also be prepared using a similar growing scheme starting from initially decoupled patches and find that despite the finite fidelity in preparing the quasi-hole state the fractional charge remains robustly quantized and hence provides a clear indicator of the topological nature of the prepared state.
We benchmark our results by comparing with a more conventional protocol based on coupling one-dimensional chains~\cite{He2017}.

\section{Model}
We study the Hofstadter-Bose-Hubbard model on a square lattice of size $L_x \times L_y$, which reads in Landau gauge 
\begin{equation}
	\begin{aligned}
		&\hat{\mathcal{H}}_{L_x\times L_y} = \\
		&\phantom{\hat{\mathcal{H}}}-J \sum_{x,y} \left(\hat{a}^{\dagger}_{x+1,y}\hat{a}_{x,y}^{\vphantom\dagger} + \mathrm{e}^{2\pi i \alpha x}\hat{a}^{\dagger}_{x,y+1}\hat{a}_{x,y}^{\vphantom\dagger} + \mathrm{H.c.}\right)\\
		&\phantom{\hat{\mathcal{H}}}+\frac{U}{2}\sum_{x,y} \hat{n}_{x,y}\left(\hat{n}_{x,y}-1\right).
	\end{aligned}
\end{equation}
Here, $\hat{a}^{(\dagger)}_{x,y}$ annihilates (creates) a boson at site $(x,y)$ and $\hat{n}_{x,y} = \hat{a}_{x,y}^{\dagger}\hat{a}_{x,y}^{\vphantom\dagger}$ is the particle number operator.
In this work, we consider strong but finite Hubbard repulsion of strength $\nicefrac{U}{J} = 8$, realistic for cold atom experiments, but also infinitely strong Hubbard repulsion, $\nicefrac{U}{J} = \infty$ (hard-core).

In our simulations, we make use of exact diagonalization (ED) for small systems and tensor network methods for larger systems.
In particular, we use the \textsc{SyTen} toolkit~\cite{HubigSyTen} to perform density-matrix renormalization group (DMRG)~\cite{White1992,Schollwoeck2005,Schollwoeck2011} simulations to explore static properties and its implementation of the time-dependent variational principle (TDVP)~\cite{Haegeman2011,Haegeman2016} for dynamical calculations.
In all our simulations we exploit the $\mathrm{U}(1)$-symmetry associated with particle number conservation and consider bond dimensions up to $\chi=1024$.

\section{Coupling two $4\times4$-patches}
We begin our discussion of patchwork growing schemes with proof-of-principle simulations of the minimal system consisting of two initially decoupled patches.
We envision two experimentally relevant approaches to grow extended FCIs.
On the one hand, we consider completely decoupled patches, which are then connected by turning on a local hopping $J_{\rm coupling}$ between the two patches, Fig.~\ref{fig:2Patches_Sketch}.
On the other hand, we study a system of $9\times4$ sites with a strong potential barrier $V_{\rm barrier}$ on the center column, splitting the system into two $4\times4$-patches, Fig.~\ref{fig:2Patches_Sketch_Vbarrier}.
Upon turning off the potential barrier, we can grow an extended FCI state in the enlarged system.

For the small systems studied in this section, we exactly diagonalize the Hamiltonian and also treat the subsequent time-evolution simulation in this manner.
Only later on we will use the matrix product state (MPS) representation, when discussing the entanglement growth upon time-evolution.

\subsection{Hopping Protocol: Turning On Local Hoppings\label{sec:HoppingProtocol}}
\begin{figure}[b]
	\centering
	\includegraphics{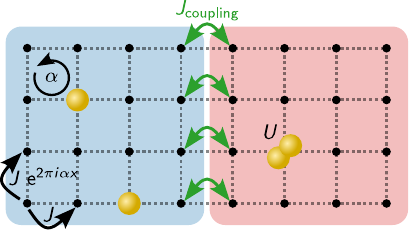}
	\caption{
		\textit{Hopping protocol}:
		Coupling two $4\times4$ Hofstadter-Bose-Hubbard patches with initially $N=2$ particles in each half of the system, indicated by the blue and red shading respectively.
		The parameter $J_{\rm coupling}$ is dynamically adjusted to connect the initially decoupled patches into one extended system.
	}
	\label{fig:2Patches_Sketch}
\end{figure}

We first discuss the ``\textit{hopping protocol}'' which takes two $4\times4$-patches and couples them by a local hopping across the edge connecting the patches, see Fig.~\ref{fig:2Patches_Sketch}.
This system is described by the Hamiltonian
\begin{equation}
	\begin{aligned}
		\hat{\mathcal{H}}_{2p}(J_{\rm coupling}) = &~\hat{\mathcal{H}}_{4\times 4}^{(1)} + \hat{\mathcal{H}}_{4\times 4}^{(2)}\\
		&~- J_{\rm coupling} \sum_{y = 1}^{4} \left( \hat{a}_{5, y}^{\dagger} \hat{a}_{4, y}^{\vphantom\dagger} + \mathrm{H.c.} \right),
	\end{aligned}
\end{equation}
where the first two terms describe the independent patches and the last term describes the coupling between the two patches.

\subsubsection{Static properties}
Before performing time-evolution simulations of the coupling process, we consider the many-body gap $\Delta$ between the ground state and the first excited state to find suitable parameters for the adiabatic protocol.
Furthermore, we confirm the topological nature of the target-state by extracting the many-body Chern number of the ground state.
To this end, we diagonalize the Hamiltonian for varying coupling strength $J_{\rm coupling}$ and flux $\alpha$.

\begin{figure*}
	\centering
	\includegraphics{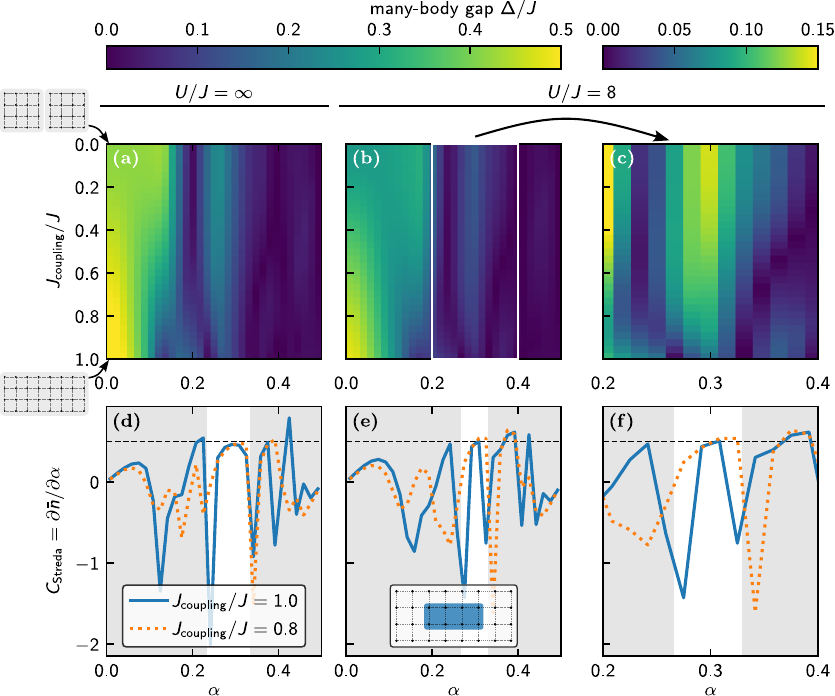}
	\caption{
		\emph{Hopping protocol}:
		\textbf{(a-c)}
		Many-body gap $\Delta$ obtained from the exact diagonalization of a system of two coupled $4\times 4$-patches for \textbf{(a)} $\nicefrac{U}{J} = \infty$ and \textbf{(b,~c)} $\nicefrac{U}{J} = 8$.
		We tune both the flux per plaquette $\alpha$ and the coupling strength $J_{\rm coupling}$ between the two patches.
		Panel \textbf{(c)} is a zoom-in of \textbf{(b)} focusing on the regime relevant to the $\nu=\nicefrac{1}{2}$ Laughlin state, characterized by the opening of the many-body gap in the decoupled patches at $\alpha\approx 0.25$.
		Note that in this regime the gap remains open for all $J_{\rm coupling}$ and therefore provides a promising path for adiabatic state preparation.
		\textbf{(d-f)}
		Many-body Chern number $C_{\rm Streda}$ in the fully coupled system obtained using St\v{r}eda's formula.
		In the gapped region discussed before we find $C_{\rm Streda}=\nicefrac{1}{2}$ as expected for the $\nu=\nicefrac{1}{2}$ Laughlin state.
		Grayed out regions are not expected to host the Laughlin state at $\nicefrac{J_{\rm coupling}}{J} \approx 1$.
        The inset in \textbf{(e)} visualizes the bulk region used for St\v{r}eda's formula.
	}
	\label{fig:2Patches_MBgap_And_Streda}
\end{figure*}

Fig.~\ref{fig:2Patches_MBgap_And_Streda} reveals a finite many-body gap around $\alpha \approx 0.3$, which adiabatically connects the decoupled ($\nicefrac{J_{\rm coupling}}{J}=0$) to the homogeneously coupled system ($\nicefrac{J_{\rm coupling}}{J}=1$).
This indicates the possibility to grow a larger Laughlin state from two decoupled Laughlin states.
In particular, we find that this behavior persists for both hard-core bosons ($\nicefrac{U}{J} = \infty$) and for finite Hubbard repulsion ($\nicefrac{U}{J} = 8$).

To confirm the topological nature of the ground state, we extract the many-body Chern number using St\v{r}eda's formula~\cite{Streda1982,Streda1982a,Repellin2020,Leonard2023a},
\begin{equation}
	C_{\rm Streda} = 2\pi \frac{\partial \bar{n}}{\partial \varphi} = \frac{\partial \bar{n}}{\partial \alpha},
\end{equation}
where $\bar{n}$ is the bulk density of the system.
We consider the central $4\times2$ sites in the fully coupled ($J_{\rm coupling}=J$) $8\times4$-system to evaluate the bulk density $\bar{n}$, and we find the Chern number in the gapped phase around $\alpha\approx 0.3$ to be $C_{\rm Streda}\approx\nicefrac{1}{2}$, see Fig.~\ref{fig:2Patches_MBgap_And_Streda}.
This agrees with the expected value for the continuum $\nu=\nicefrac{1}{2}$ Laughlin state, so that we conclude that the homogeneously coupled system at $J_{\rm coupling} = J$ indeed hosts a lattice analog of this topologically ordered target-state.

\subsubsection{Time-evolution}
Motivated by these findings, we consider a simple preparation protocol, where the hopping amplitude between the two patches is increased linearly over a finite preparation time $T$,
\begin{equation}
	J_{\rm coupling}(t) = \frac{t}{T}~J,
\end{equation}
while the magnetic flux per plaquette is kept fixed, \mbox{$\alpha=0.3$}.
Integrating the time-dependent Schr\"odinger equation, we calculate the time-evolution of the initial product state
\begin{equation}
	\ket{\Psi_{\rm initial}} = \ket{\Psi(t = 0)} = \ket{\Psi_{0}^{(1)}} \otimes \ket{\Psi_{0}^{(2)}},
\end{equation}
where $\ket{\Psi_{0}^{(k)}}$ is the ground state in the initially decoupled $4\times 4$-patches labeled by $k$.
Our target-state $\ket{\Psi_{\rm target}}$ is the ground state of the large $8\times4$-system.
We determine the target-state fidelity of the time-evolved state $\ket{\Psi(t)}$, which is defined as
\begin{equation}
	\mathcal{F}(t) = |\braket{\Psi_{\rm target} | \Psi(t)}|^2.
\end{equation}

\begin{figure}
	\centering
	\includegraphics{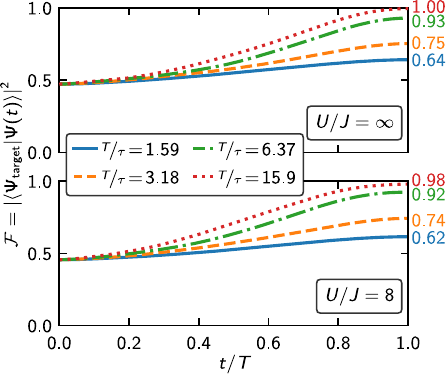}
	\caption{
		\emph{Hopping protocol}:
		Target-state fidelity of the time-evolved state $\ket{\Psi(t)}$ for various preparation times $T$, in units of $\tau = \frac{2\pi\hbar}{J}$, upon linearly coupling two patches.
	}
	\label{fig:2Patches_Coupling}
\end{figure}

Already without any further optimization of the preparation path, we find a final fidelity \mbox{$\mathcal{F}(T) > 0.9$} for preparation times $\nicefrac{T}{\tau} \gtrsim 6$, see Fig.~\ref{fig:2Patches_Coupling}, where $\tau = \frac{2\pi\hbar}{J}$ is the characteristic tunneling time and we use natural units where $\hbar = 1$.
In our simulations, we considered different preparation times $\nicefrac{T}{\tau} = 1.59, 3.18, 6.37, 15.9$ (corresponding to $TJ = 10, 20, 40, 100$, respectively) and a path along which the minimal many-body gap is \mbox{$\Delta_{\rm min} \approx 0.09 J$}.
Note however, that for most of the ramp the many-body gap is significantly larger, so that an improved, potentially non-linear, ramp of the coupling is expected to result in substantially shorter preparation times with comparable target-state overlap.

\begin{figure*}
	\centering
	\includegraphics{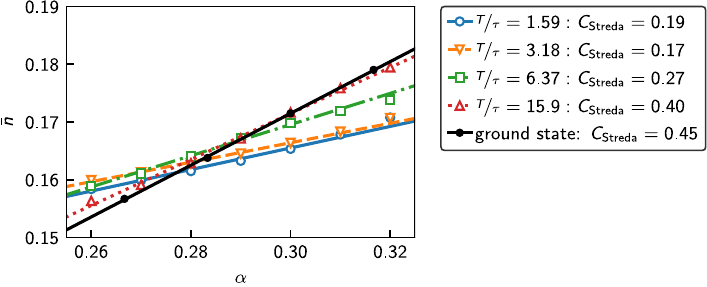}
	\caption{
		\emph{Hopping protocol}:
		Bulk density $\bar{n}$ as a function of flux per plaquette $\alpha$ for the ground state (black) and the final state of the time-evolution simulations for different times (colored).
		Using a linear fit according to Streda's formula, we obtain estimates for the Chern number consistent with the $\nicefrac{1}{2}$-Laughlin state for a sufficiently slow ramp.
		Data is given for $\nicefrac{U}{J}=\infty$.
	}
	\label{fig:2Patches_Streda}
\end{figure*}

We confirm the topological nature of the prepared state at the end of the time-evolution by performing similar calculations for different values of the flux per plaquette $\alpha$ and extracting the Chern number $C_{\rm Streda}$ using St\v{r}eda's formula, see Fig.~\ref{fig:2Patches_Streda}.
For a sufficiently slow ramp, \mbox{$\nicefrac{T}{\tau}=15.9$}, we find a value of $C_{\rm Streda}$ consistent with the $\nicefrac{1}{2}$-Laughlin state expected close to $\alpha=0.3$.
We attribute the deviation from the ground state value to the finite population of excited state upon time-evolution.
The slight discrepancy between the ground state's value and the expected value $C=\nicefrac{1}{2}$ can be attributed to finite size effects.

\subsubsection{Entanglement growth and adiabaticity}
\begin{figure}
	\centering
	\includegraphics{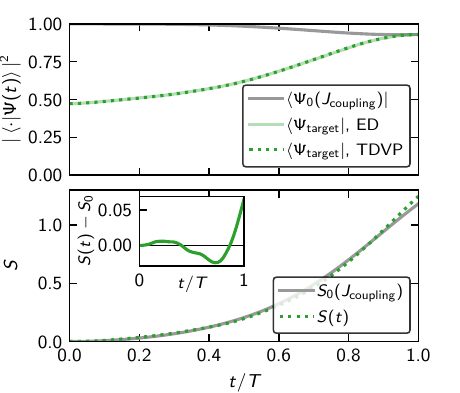}
	\caption{
		\textit{Hopping rotocol}:
		In the upper panel, the squared overlaps of the state obtained by performing the time-evolution via TDVP for an MPS (green dotted line) and exact diagonalization (light green line) are in excellent agreement.
		We find a large overlap of the time-evolved state with the instantaneous ground state at all times (gray line).
		Furthermore, we observe the expected entanglement growth (lower panel) as the two patches are coupled both for the time-evolved state (green) and the instantaneous ground state (gray).
		Inset: The excess entropy in the time-evolved state at $\nicefrac{t}{T}=1$ is attributed to a small population of excited states.
		Data is given for $\nicefrac{U}{J} = \infty$, $\nicefrac{T}{\tau}=6.37$ and $N_{\rm Trotter}=40$.
	}
	\label{fig:EntanglementGrowth}
\end{figure}

While the initial state is a product state for which the entanglement entropy between the patches vanishes, $S(t=0) = 0$, the target-state is clearly an entangled state, $S(t = T) > 0$.
We will now probe this behavior using MPS simulations, allowing for a particularly simple extraction of the bipartite entanglement entropy between spatial subsystems.
To this end, we perform time-evolution simulations using the time-dependent variational principle (TDVP) method for MPS and track the entanglement entropy $S(t)$ between the two $4\times4$-subsystems.
We perform a Trotterization of the time-evolution operator,
\begin{equation}
	\hat{\mathcal{U}} = \mathcal{T} \mathrm{e}^{-i\int_0^T \mathrm{d}t~\hat{\mathcal{H}}_{2p}(t) } \approx \prod_{n = 0}^{N_{\rm Trotter}} \mathrm{e}^{-i \delta t \hat{\mathcal{H}}_{2p}(t_{n})},
\end{equation}
where $\mathcal{T}$ is the time-ordering operator, $\delta t = \nicefrac{T}{N_{\rm Trotter}}$, $t_n = n\delta t$, and
\begin{equation}
	\hat{\mathcal{H}}_{2p}(t) = \hat{\mathcal{H}}_{2p}(J_{\rm coupling}(t)),
\end{equation}
and consider a preparation time of $\nicefrac{T}{\tau} = 6.36$, for which we achieve convergence in the Trotter step size for \mbox{$N_{\rm Trotter} = 40$}, which also manifests in excellent agreement of the target-state fidelity for the ED and TDVP time-evolution simulations, see Fig.~\ref{fig:EntanglementGrowth}.

In our time-evolution simulation, we find the expected growth of entanglement as the patches are connected, see Fig.~\ref{fig:EntanglementGrowth}.
In particular, the final entanglement entropy is slightly larger than in the instantaneous ground state of the $8\times4$-system, $S_0(J_{\rm coupling})$, see also the inset in Fig.~\ref{fig:EntanglementGrowth}.
We attribute this to a slight deviation from a perfectly adiabatic protocol, resulting in a population of excited states as we connect the patches, which is also visible in the slight decrease of the overlap with the instantaneous ground state towards the end of the preparation protocol.
Slowing down the ramp in this regime (while speeding it up at early times) might give access to more accurate and, more importantly, faster preparation protocols.

\subsection{Barrier Protocol: Turning Off a Barrier}
Reassured by this proof-of-principle, we next turn to an alternative preparation protocol where we propose to start from a system of size $9\times 4$, initially split into two halves by a tunable potential barrier.
\begin{figure}
	\centering
	\includegraphics{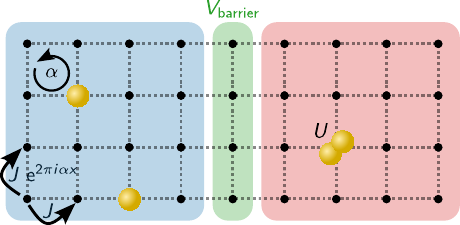}
	\caption{
		\textit{Barrier protocol}:
		Coupling two $4\times4$ Hofstadter-Bose-Hubbard patches with $N=2$ particles in each half of the system, indicated by the blue and red shading respectively.
		The patches are separated by a local potential barrier (green region), which is dynamically lowered to connect the patches into one extended $9\times4$ system.
	}
	\label{fig:2Patches_Sketch_Vbarrier}
\end{figure}
The Hamiltonian for such a system, realizing what we call the ``\emph{barrier protocol}'', reads
\begin{equation}
	\hat{\mathcal{H}}_{2p}(V_{\rm barrier}) = \hat{\mathcal{H}}_{9\times 4} + V_{\rm barrier} \sum_{y = 1}^{4} \hat{n}_{5, y},
\end{equation}
where $\nicefrac{V_{\rm barrier}}{J}$ is the height of the potential barrier, see Fig.~\ref{fig:2Patches_Sketch_Vbarrier}.
Such an approach is experimentally most realistic in quantum gas microscopes, where digital micromirror devices (DMDs)~\cite{Zupancic2016} or spatial light modulators (SLMs)~\cite{Gaunt2012} can be used to create such barriers.

\subsubsection{Static properties}

\begin{figure*}
	\centering
	\includegraphics{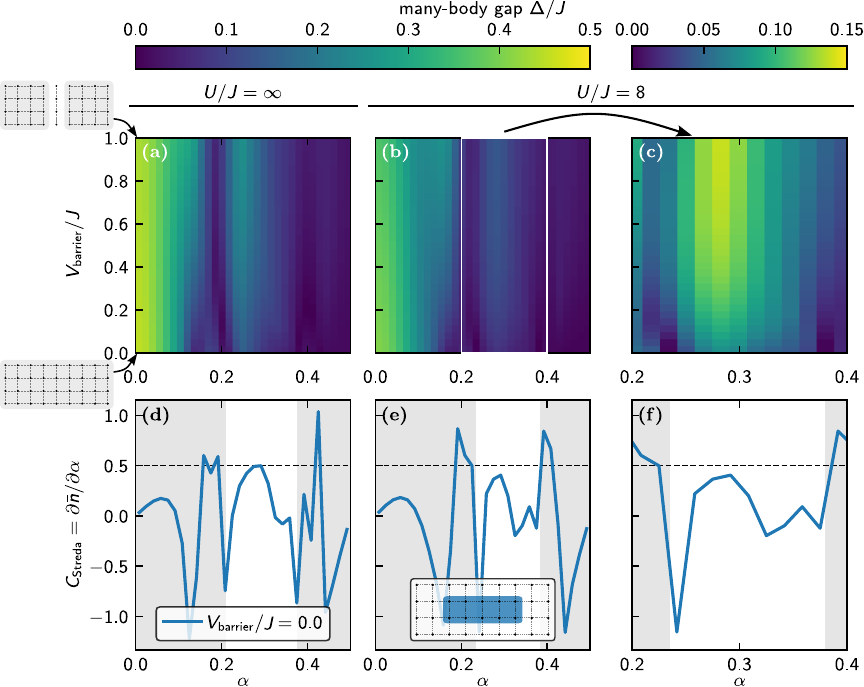}
	\caption{
		\emph{Barrier protocol}:
		\textbf{(a-c)}
		Many-body gap $\Delta$ obtained from the exact diagonalization of a system of two coupled $4\times 4$-patches for \textbf{(a)} $\nicefrac{U}{J} = \infty$ and \textbf{(b,~c)} $\nicefrac{U}{J} = 8$.
		We tune both the flux per plaquette $\alpha$ and the height of the potential barrier $V_{\rm barrier}$ between the two patches.
		Panel \textbf{(c)} is a zoom-in of \textbf{(b)} focusing on the regime relevant to the $\nu=\nicefrac{1}{2}$ Laughlin state, characterized by the opening of the many-body gap in the decoupled patches at $\alpha\approx 0.25$.
		Note that in this regime the gap remains open for all $V_{\rm barrier}$ and therefore provides a promising path for adiabatic state preparation.
		\textbf{(d-f)}
		Many-body Chern number $C_{\rm Streda}$ of the target-state ($V_{\rm barrier}=0$) obtained using St\v{r}eda's formula.
		In the gapped region discussed before we find $C_{\rm Streda}\approx\nicefrac{1}{2}$ as expected for the $\nu=\nicefrac{1}{2}$ Laughlin state.
		Grayed out regions are not expected to host the Laughlin state at $\nicefrac{V_{\rm barrier}}{J} \approx 0$.
        The inset in \textbf{(e)} visualizes the bulk region used for St\v{r}eda's formula.
	}
	\label{fig:2PatchesBarrier_MBgap}
\end{figure*}

Again, we diagonalize the Hamiltonian for hard-core bosons ($\nicefrac{U}{J}=\infty$) and finite Hubbard repulsion \mbox{($\nicefrac{U}{J}=8$)} and find a path with a finite excitation gap from the decoupled regime to the fully coupled system without any barrier, see Fig.~\ref{fig:2PatchesBarrier_MBgap}.
In particular, we note that already relatively weak local potentials, $\nicefrac{V_{\rm barrier}}{J} \approx 1$, provide a sufficient barrier to split the system into two parts.
Note however, that while the excitation gap is almost constant above $\nicefrac{V_{\rm barrier}}{J} = 0.5$, the microscopic structure might still differ slightly from the completely decoupled limit.
We will encounter this behavior below in the analysis of our time-evolution simulations.

Calculating the many-body Chern number of the target-state using St\v{r}eda's formula as above, we again find evidence for a topologically non-trivial state in the gapped regime of the $9\times4$-system, see Fig.~\ref{fig:2PatchesBarrier_MBgap}.
Furthermore, the extracted Chern number $C_{\rm Streda} \approx \nicefrac{1}{2}$ is consistent with the expectation for the $\nicefrac{1}{2}$-Laughlin state, despite significant deviations from the exactly quantized value attributed to the small size ($5\times2$) of the central bulk region used.

\subsubsection{Time-evolution: Linear ramp}

\begin{figure}[t]
	\centering
	\includegraphics{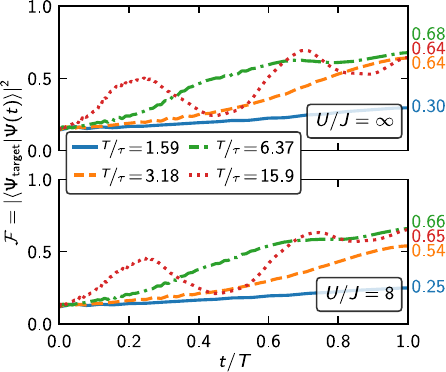}
	\caption{
		\emph{Barrier protocol}:
		Target-state fidelity of the time-evolved state $\ket{\Psi(t)}$ for varying preparation times $T$, in units of $\tau = \frac{2\pi\hbar}{J}$, upon linearly turning off the potential barrier, starting from $V_{\rm barrier}(t=0)=V_0=5J$.
		The target-state overlaps are significantly smaller than for the hopping protocol discussed above.
	}
	\label{fig:2Patches_Barrier}
\end{figure}

Performing a similar time-evolution protocol as above, we again calculate the overlap of the time-evolved state with the target-state for a time-dependent potential barrier of height
\begin{equation}
	V_{\rm barrier}(t) = V_0 \left(1 - \frac{t}{T}\right),
\end{equation}
where we fix the initial potential barrier $V_0=5J$ and the flux per plaquette $\alpha=0.3$ in our simulations.
The initial state is taken to be a product state of two $4\times4$-patches, corresponding to the ground state in the limit $V_{\rm barrier}\to\infty$.

As for the hopping protocol, we find that the target-state fidelity is already substantial ($\mathcal{F}(T) \approx 0.6$) for relatively short ramp times ($\nicefrac{T}{\tau} \approx 3$), see Fig.~\ref{fig:2Patches_Barrier}.
However, the simple linear ramp of the barrier does in general not give as good target-state overlap as a direct coupling of the two patches discussed in Section~\ref{sec:HoppingProtocol}.

Furthermore, we observe an oscillatory behavior of the target-state fidelity especially for slow ramps.
We attribute this to the fact that while the initial product state is an approximate eigenstate of the Hamiltonian in the presence of a barrier of finite height $V_0$, this approximation becomes exact only in the limit $V_0 \to \infty$.
I.e. there is always a small occupation of excited states present.

Moreover we conclude that quickly ramping down the barrier early on during the time-evolution, where the initial state is still an approximate eigenstate of the time-dependent Hamiltonian, might not decrease the intermediate fidelity significantly.
In contrast, spending more time on the later part of the evolution, where the excitation gap is small and adiabaticity is more difficult to reach, might prove useful in reaching higher overall target-state fidelities.

\subsubsection{Time-evolution: Two-step ramp}

\begin{figure}
	\centering
	\includegraphics{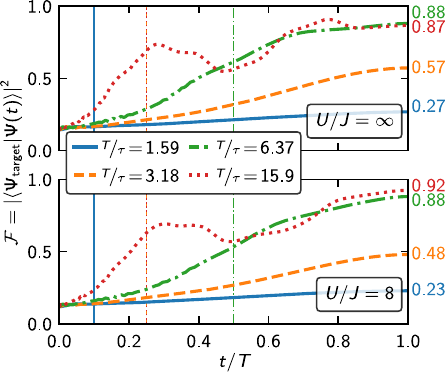}
	\caption{
		\emph{Two-step barrier protocol}:
		Target-state fidelity of the time-evolved state for varying preparation times $T$, in units of $\tau = \frac{2\pi\hbar}{J}$,  upon turning off the potential barrier in two linear steps, starting from $V_{\rm barrier}(t=0)=V_0=5J$.
		We vary the intermediate time $T_1$ at which the piecewise linear ramp passes through $V_{\rm barrier}(T_1) = J$ (indicated by vertical lines).
	}
	\label{fig:2Patches_Barrier:TwoStep}
\end{figure}

Based on this observation, we modify the protocol by splitting up the turn-off of the potential barrier into two parts.
Specifically, we linearly reduce the barrier from $V_{\rm barrier}(0) =V_0 = 5J$ to $V_{\rm barrier}(T_1) = J$ over a time $T_1 < T$.
Then, we slowly turn off the barrier completely during the remaining time $T-T_1$.
We can write this piecewise linear function as
\begin{equation}
	V_{\rm barrier}(t) = 
	\begin{cases}
		\frac{T_1 - t}{T_1}~V_0 + \frac{t}{T_1}~J & \text{ for } t \leq T_1, \\
		\frac{T - t}{T-T_1}~J & \text{ for } t \geq T_1.
	\end{cases}
\end{equation}

The two parts of the modified ramp can be intuitively understood as follows:
The first part, $t \leq T_1$, relatively quickly takes the system from the essentially decoupled regime to a point closer to the coupled case.
The non-trivial evolution from an (approximate) product state of decoupled patches to the entangled state in the fully coupled limit is performed in the second part, $t \geq T_1$.
Instead of performing the entire time-evolution at the same ramp speed, the region of smaller many-body gaps (i.e. small $\nicefrac{V_{\rm barrier}}{J}$) is traversed slower compared to the simple linear ramp.

As expected, the modified ramp performs better at reaching high fidelity with the target-state at fixed total ramp time if the time $T_1$ is chosen sufficiently large, see Fig.~\ref{fig:2Patches_Barrier:TwoStep}.
Here, we only perform some exemplifying time-evolution simulations to show the general advantage of piecewise linear ramps over a simple linear ramp, varying the ratio $T_1/T$ between $0.1$ and $0.5$, which could be optimized for different total ramp times individually in the future.
We believe further improvement to be possible by using more complicated ramp profiles which can be optimized using, for example, machine learning techniques.

\subsection{Intermediate Conclusions}
By successfully growing an extended Laughlin state from two patches with high fidelity, we provided a numerical proof-of-principle of simple patchwork preparation schemes.
Specifically, we conclude that it is in principle possible to prepare an extended Laughlin state in cold atom experiments by connecting two originally independent patches of $2$ particles on $4\times 4$ sites.

\begin{figure*}
	\centering
	\includegraphics{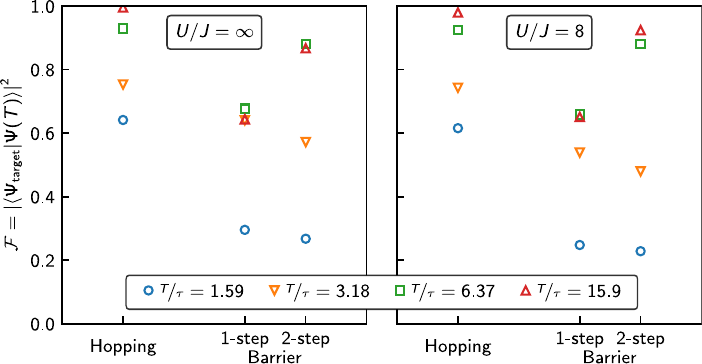}
	\caption{
		Fidelity of the final time-evolved state $\ket{\Psi(t=T)}$ for the preparation protocols discussed above.
		We find good performance of all protocols for sufficiently slow ramps with the hopping protocol out-performing both the one- and two-step barrier protocols.
	}
	\label{fig:2Patches_Comparison}
\end{figure*}

In our simulations we considered two different protocols called the ``hopping protocol'' and the ``barrier protocol'', respectively, both allowing for target-state fidelities of $\mathcal{F}(T) \gtrsim 0.85$ for preparation times on the order $\nicefrac{T}{\tau} \gtrsim 6$, see Fig.~\ref{fig:2Patches_Comparison}.
In general, the hopping protocol achieved the highest target-state fidelity for a given ramp time $T$, however the two-step barrier protocol yields comparable results for $\nicefrac{T}{\tau} \gtrsim 6$.
Further optimization of the ramp parameters as well as more sophisticated, potentially non-linear, ramps might allow for even higher fidelities for short preparation times.

\section{Super-Chains: Coupling more patches}
Elongated systems provide interesting new insights involving the edge theory of FCI states allowing, for example, for measurements of the central charge~\cite{Palm2022}.
Despite their finite size in the short direction, they were found to exhibit gapless chiral edge modes for sufficiently long systems.
Here, we propose to couple more than two patches - hosting two particles each - to realize such elongated systems, see Fig.~\ref{fig:Fig1}\textbf{(a)}.
We call a collection of patches in a chain-like structure a ``super-chain'', which is described by the Hamiltonian
\begin{equation}
	\begin{aligned}
		\hat{\mathcal{H}}_{\rm chain}&\left(\left\{J_{\rm coupling}^{(k)}\right\}\right) = \sum_{k = 1}^{N_{\rm patches}} \hat{\mathcal{H}}_{4\times 4}^{(k)} \\
		&- \sum_{k=1}^{N_{\rm patches}-1} J_{\rm coupling}^{(k)} \sum_{y = 1}^{4} \left( \hat{a}_{4k+1, y}^{\dagger} \hat{a}_{4k, y}^{\vphantom\dagger} + \mathrm{H.c.} \right),
	\end{aligned}
	\label{eq:HamiltonianSuperchain}
\end{equation}
where the couplings $J_{\rm coupling}^{(k)}$ between the different patches may be adjusted independently.
As systems of $N_{\rm patches} = 3,4,5$ patches become too large to be treated with exact diagonalization, we turn to simulations based on MPS to first find the low-lying states of the system and afterwards perform time-evolution.

\subsection{Static Properties}
We start with a system of homogeneous inter-patch couplings, $J_{\rm coupling}^{(1)} = \hdots = J_{\rm coupling}^{(N_{\rm patches}-1)} = J_{\rm coupling}$, and restrict our analysis to the case $\nicefrac{U}{J}=\infty$ for numerical convenience.
Given our earlier findings we do not expect qualitative changes in the case of finite Hubbard repulsion.
Using DMRG searches for the ground state and the first excited state, we calculate the many-body gap $\Delta$.
Furthermore, we extract the entanglement entropy $S^{(1)}$ between the first patch and the rest of the system from the ground state MPS.

\begin{figure*}
	\centering
	\includegraphics{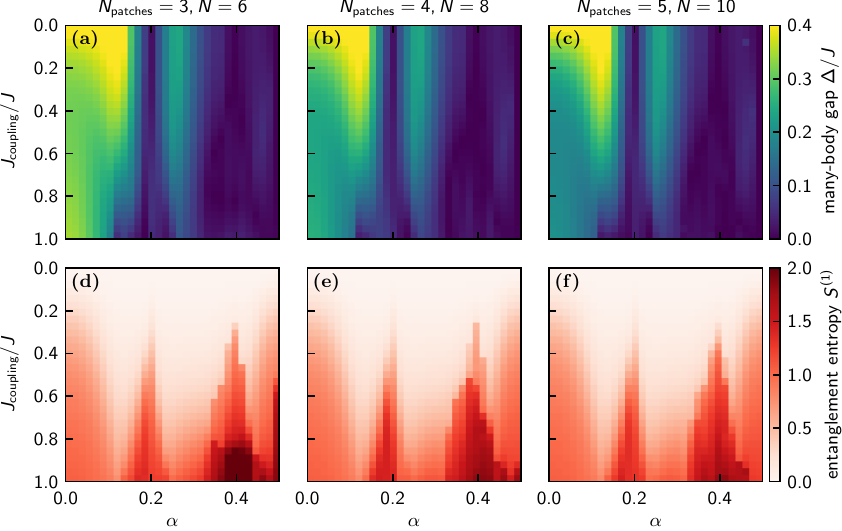}
	\caption{
		\textit{Super-chains}:
		Coupling $N_{\rm patches} = 3, 4$ or $5$ patches into a super-chain results in essentially unchanged behavior for the many-body gap $\nicefrac{\Delta}{J}$ \textbf{(a-c)} and the entanglement entropy between the first patch and the remaining system \textbf{(d-f)}.
		While the gaps are quantitatively smaller than for the case of $N_{\rm patches}=2$ (see Fig.~\ref{fig:2Patches_MBgap_And_Streda}), the relatively large gap related to the Laughlin state on a single patch ($J_{coupling}/J =0$) remains clearly visible in all cases, see also Fig.~\ref{fig:ScalingMBGap} in the Appendix, and allows for an adiabatic state preparation.
	}
	\label{fig:SuperChains_MBgap}
\end{figure*}

Irrespective of the number of patches, we find a gapped state at $\alpha \approx 0.3$ connected to the Laughlin state on a single patch, see Fig.~\ref{fig:SuperChains_MBgap}.
The gap around $\alpha\approx0.3$ persists for all values of the coupling between the patches, so that we believe an adiabatic preparation of elongated systems from smaller patches to be realistic.
At the same time, the entanglement between the first patch and the remaining system remains relatively small compared to the gapless states at larger and smaller flux per plaquette $\alpha$.
We interpret this as additional indication for the Laughlin state extending to large systems, as the Laughlin state is gapped, exhibiting only area-law entanglement~\cite{Eisert2010,Cirac2021}.

\subsection{Time-Evolution}
As for the two-patch systems above, we perform time-evolution simulations starting from a product state
\begin{equation}
	\ket{\Psi_{\rm initial}} = \bigotimes_{k=1}^{N_{\rm patches}} \ket{\Psi_0^{(k)}}.
\end{equation}
We employ the TDVP method to evolve the MPS representing the initial state.
We evolve the state with the time-dependent form of the Hamiltonian in Eq.~\ref{eq:HamiltonianSuperchain}, where the time-dependence enters via the hopping amplitudes $J_{\rm coupling}^{(k)}(t)$.
In all our simulations we keep the flux per plaquette fixed to $\alpha=0.3$.

\subsubsection{Coupling all patches at once}
\begin{figure*}
	\centering
	\includegraphics{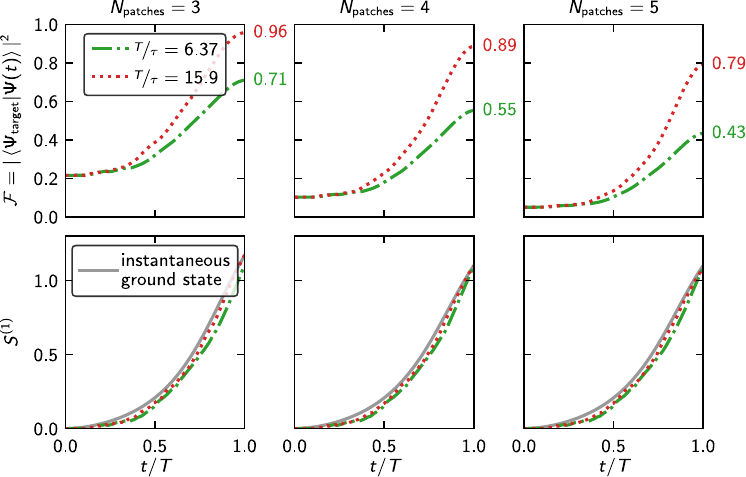}
	\caption{
		\textit{Super-chains}:
		Target-state fidelity $\mathcal{F}$ (top) and entanglement entropy $S^{(1)}$ (bottom) of the first patch upon linearly coupling all patches simultaneously.
	}
	\label{fig:SuperChains_Overlaps}
\end{figure*}

In a first attempt to grow extended super-chains, we couple all patches simultaneously.
We find that this simple protocol provides target-state fidelities of \mbox{$\mathcal{F}(T) \gtrsim 0.4~(0.8)$} for $\nicefrac{T}{\tau}=6.37~(15.9)$, see Fig.~\ref{fig:SuperChains_Overlaps}.
As before, we find that the most significant build-up of target-state fidelity as well as entanglement between the patches occurs during the second half of the preparation time, so that improving the ramp might increase the overall target-state fidelity.

As expected, we find the target-state fidelity at the end of the ramp to decrease as more patches are added.
We address this decrease and its connection to the scaling of the many-body gap in Appendix~\ref{Appendix:ScalingGap}.
While the decrease limits the scalability beyond a certain system size, we are confident that extended systems relevant for studies of edge properties like the central charge are within reach.
Nevertheless, below we will discuss alternative protocols to grow extended systems from originally decoupled patches to see whether we can reach even higher target-state fidelities in large systems.

\subsubsection{Iteratively coupling pairs of patches}
For an even number of patches it might be favorable to first couple them in pairs of two patches before coupling the pairs among each other in order to optimally exploit the finite-size gaps.
Here, we investigate this concept for a set of $N_{\rm patches} = 4$ patches.
We split the preparation time $T$ in equal intervals, such that we first couple two patches each to form two pairs ($t \leq \nicefrac{T}{2}$) and then connect the pairs ($t > \nicefrac{T}{2}$).

\begin{figure}[b]
	\centering
	\includegraphics{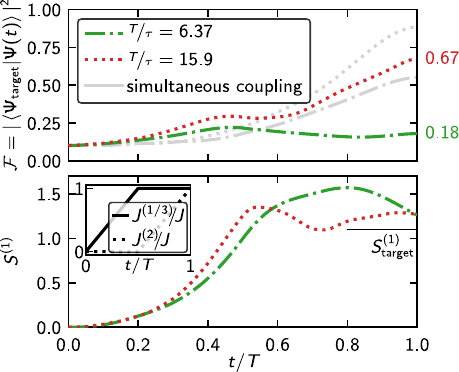}
	\caption{
		\textit{Super-chains}:
		Target-state fidelity $\mathcal{F}$ and entanglement entropy $S^{(1)}$ of the first patch upon first growing two pairs of patches ($\nicefrac{t}{T} < 0.5$) and afterwards connecting the pairs ($\nicefrac{t}{T} \geq 0.5$).
		The gray lines in the upper panel indicate the overlap achieved when all hoppings are ramped up simultaneously.
		The inset in the lower panel visualizes the different intra- ($J^{(1)}=J^{(3)}$) and inter-pair ($J^{(2)}$) coupling strengths.
	}
	\label{fig:SuperChains_Overlaps:TwoThenTwo}
\end{figure}

For $\nicefrac{T}{\tau} = 6.37$ and $15.9$ we find that this protocol yields target-state fidelities which are significantly \emph{lower} than those from a simultaneous coupling of all patches, see Fig.~\ref{fig:SuperChains_Overlaps:TwoThenTwo}.
Furthermore, the time-evolved state exhibits larger entanglement entropy than the target-state, clearly showing the enhanced population of excited states.
We attribute this to the relatively fast ramp in the first half of the preparation, when connecting pairs of two patches, recalling that preparation times $\nicefrac{T}{\tau}\gtrsim 6$ were needed to reach target-state fidelities above $90\%$ there, see Fig.~\ref{fig:2Patches_Coupling}.
Optimizing the relative length of both intervals might result in higher overlaps at the end of the complete ramp.

\subsubsection{Attaching one patch at a time}
We conclude our analysis of the super-chain setup by studying a protocol where the patches are attached consecutively.
In particular, we linearly ramp up the couplings $J_{\rm coupling}^{(k)}$, $k = 1, \hdots, N_{\rm patches}-1$, between the patches one after the other:
\begin{widetext}
	\begin{equation}
		\frac{J_{\rm coupling}^{(k)}(t)}{J} = \begin{cases}
			0 & \text{ for } \frac{t}{T} < \frac{k-1}{N_{\rm patches}-1}\\
			\frac{t}{T}~\left(N_{\rm patches}-1\right) - (k-1) & \text{ for } \frac{k-1}{N_{\rm patches}-1} \leq \frac{t}{T} \leq \frac{k}{N_{\rm patches}-1}\\
			1 & \text{ for } \frac{t}{T} > \frac{k}{N_{\rm patches}-1}
		\end{cases}.
	\end{equation}
\end{widetext}

\begin{figure*}
	\centering
	\includegraphics{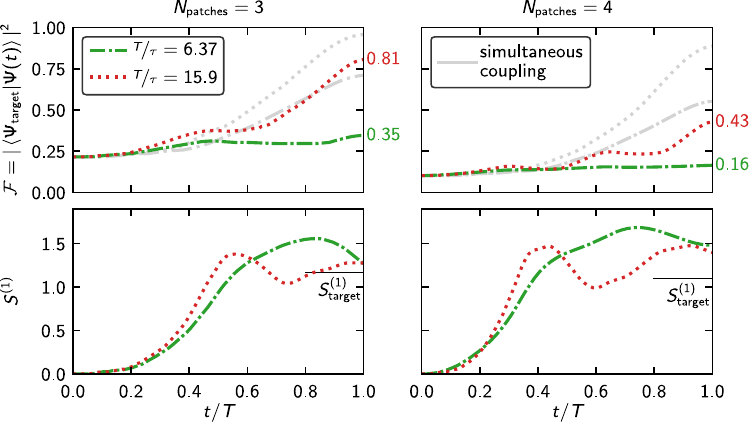}
	\caption{
		\textit{Super-chains}:
		Target-state fidelity $\mathcal{F}$ (top) and entanglement entropy $S^{(1)}$ (bottom) of the first patch upon consecutively connecting the patches.
		The gray lines in the upper row indicate the overlap achieved when all hoppings are ramped up simultaneously.
	}
	\label{fig:SuperChains_Overlaps:OneAtATime}
\end{figure*}

As before, we calculate the target-state fidelity as a function of time as well as the entanglement entropy between the first patch and the rest of the system.
While we find large target-state overlaps for the slowest ramps considered here ($\nicefrac{T}{\tau} = 15.9$), we do not find as good overlaps for a faster ramp, see Fig.~\ref{fig:SuperChains_Overlaps:OneAtATime}.
Furthermore, we find a significant increase in the entanglement entropy compared to that of the target-state.
Increasing the system size, i.e. the number of patches, further enhances these effects.
We attribute the enhanced excitation to higher energy states to the reduced time available for attaching each individual patch.
Adjusting the couplings and in particular the time spent on each connection, one might end up with better target-state overlaps, especially when optimized on a case-by-case basis.

\subsection{Intermediate Conclusions}
Based on these results, we conclude that it is in principle possible to grow extended Laughlin states in long chains.
In particular, we believe this approach to be scalable to system sizes relevant for experimental measurements of the central charge.
In our analysis it turned out most promising to simultaneously couple all patches, as this allows for relatively short overall preparation times while still changing the local couplings sufficiently slowly.
For such a protocol we achieved target-state fidelities of $\mathcal{F} \gtrsim 0.8$ for up to five patches and $\nicefrac{T}{\tau}=15.9$.

In contrast, first coupling pairs of patches and afterwards connecting the pairs did not reach similar fidelities in our analysis.
Similarly, we did not find it advantageous to couple the patches consecutively.
We note, however, that such protocols might show their full potential only once even longer systems are considered and the coupling parameters are optimized for the specific case at hand.

\section{Large square: Coupling four patches}
As a last case study, we address the question whether it is possible to grow large squares by coupling more patches.
To this end, we focus on a barrier protocol, where a $9\times 9$-system of eight particles is split into four $4\times4$-patches by a cross-shaped potential, see Fig.~\ref{fig:4PatchesSquare_MBgap}.
This system is described by the Hamiltonian
\begin{equation}
	\begin{aligned}
		&\hat{\mathcal{H}}_{\rm square}(V_{\rm barrier})\\
		&~~~= \hat{\mathcal{H}}_{9\times 9}
		+ V_{\rm barrier} \left(\sum_{y = 1}^{9} \hat{n}_{5, y} + \sum_{x = 1}^{9} \hat{n}_{x, 5} - \hat{n}_{5,5}\right).
	\end{aligned}
\end{equation}
We remark in passing that it is also possible to add an additional local potential to directly imprint a quasi-hole during the preparation of the state, as discussed later in this section.

\subsection{Static Properties}
\begin{figure*}
	\centering
	\includegraphics{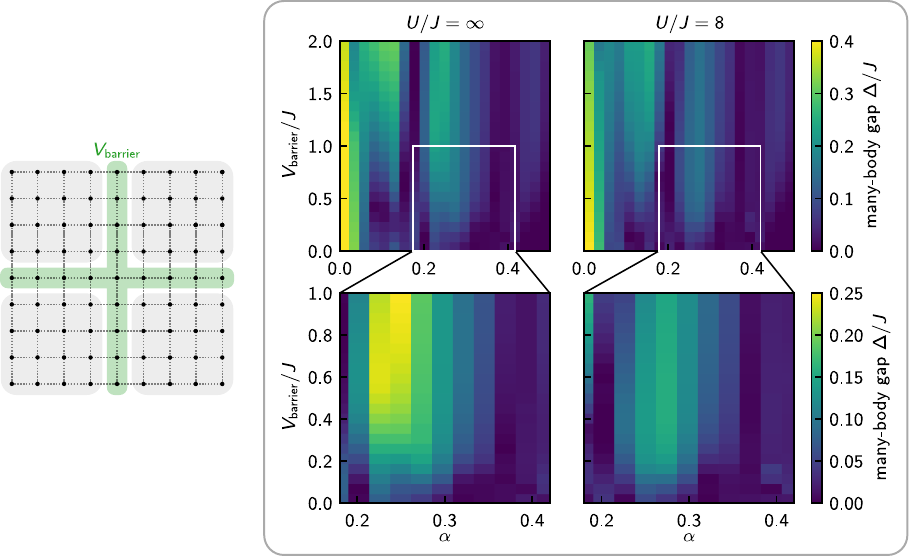}
	\caption{
		\emph{Large square}: 
		Many-body gap $\Delta$ obtained using DMRG simulations of a system of four coupled $4\times 4$-patches (sketched on the left).
		Panels in the lower row show a zoomed-in version of data in the upper row.
		We tune both the flux per plaquette $\alpha$ and the barrier height $V_{\rm barrier}$ between the patches.
		The gap of the Laughlin state at $\alpha\approx 0.25$ remains finite in all cases, even though it significantly decreases as the system gets larger.
	}
	\label{fig:4PatchesSquare_MBgap}
\end{figure*}

Once again, we perform DMRG simulations to determine the ground state and the lowest excited state, and determine the many-body gap for the case of hard-core bosons ($\nicefrac{U}{J}=\infty$) and finite Hubbard repulsion ($\nicefrac{U}{J}=8$), where in the latter case we truncate the local Hilbert space to at most four bosons per site ($N_{\rm max}=4$).
We again find an adiabatic connection from a product state of four Laughlin states to a large Laughlin state in the homogeneously coupled system, see Fig.~\ref{fig:4PatchesSquare_MBgap}.
Note, however, that close to the homogeneously coupled limit, $V_{\rm barrier}= 0$, the many-body gap is reduced as a result of the extended system size.
Nevertheless, we expect an adiabatic growing scheme for the Laughlin state in the bulk to be applicable as the low-energy excitations are believed to reside at the edge.

\subsection{Time-Evolution}
For our time-evolution simulations, we again start from a product state of four $2$-particle Laughlin states,
\begin{equation}
	\ket{\Psi_{\rm initial}} = \bigotimes_{k=1}^{4} \ket{\Psi_0^{(k)}},
\end{equation}
and use the TDVP method to simulate a preparation protocol where we linearly reduce the height of the potential barrier,
\begin{equation}
	V_{\rm barrier}(t) = V_0 \left(1 - \frac{t}{T}\right).
\end{equation}
In particular, we consider the case $\nicefrac{V_0}{J} = 1$, $\alpha=0.25$, and $\nicefrac{T}{\tau}=6.37$ and $15.9$.
We restrict our time-evolution simulations to the hard-core bosonic case, $\nicefrac{U}{J}=\infty$.

\begin{figure}
	\centering
	\includegraphics{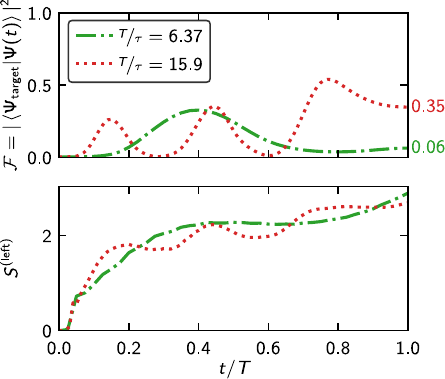}
	\caption{
		\textit{Large square}: Target-state fidelity $\mathcal{F}$ and entanglement entropy $S^{(\rm left)}$ of the left two patches for a linearly decreasing barrier height $\nicefrac{V_{\rm barrier}}{J} = 1-\nicefrac{t}{T}$ at constant flux per plaquette $\alpha=0.25$.
		Similar to the barrier protocol for only two patches (Fig.~\ref{fig:2Patches_Barrier}) we find some oscillatory behavior.
		The target-state fidelity reached at the end of the preparation protocol might be improved further by modifying the ramp similar to the barrier protocol discussed above.
	}
	\label{fig:Supersquare:Overlap}
\end{figure}

During our simulations, we monitor the target-state fidelity $\mathcal{F}(t)$ along the preparation path as well as the entanglement entropy between the left two patches and the rest of the system, see Fig.~\ref{fig:Supersquare:Overlap}.
For a sufficiently slow ramp ($\nicefrac{T}{\tau}=15.9$), we find a target-state fidelity of $\mathcal{F}(T) = 0.35$ at the end of the ramp.
Furthermore, we find a oscillatory behavior of the fidelity as for the two-patch barrier protocol discussed above.
Based on these results, it might be favorable to perform the slow ramp-down of the barrier only up to the point where the target-state fidelity is maximized, and to quickly turn off the barrier afterwards to reach the target Hamiltonian.
In conclusion, we are confident that it is possible to grow extended Laughlin states in the square geometry with sufficiently high fidelity.

\subsection{Growing a Quasi-Hole State}
As already mentioned before, a cross-shaped repulsive potential at the center of the system can be used to pin a quasi-hole emerging from the $\nu=\nicefrac{1}{2}$ Laughlin state~\cite{Raciunas2018,Macaluso2020,Wang2022}.
Next, we will first confirm that this quasi-hole in fact carries fractional charge $q_{\rm qh} = -\nicefrac{1}{2}$ as expected~\cite{Laughlin1983} and propose to use the full counting statistics of the charge to confirm this in experiments.
Afterwards, we provide numerical evidence that it is indeed possible to prepare this quasi-hole state using a patchwork approach as before and confirm the robustness of the charge fractionalization.

\subsubsection{Static properties}
\begin{figure*}
	\centering
	\includegraphics{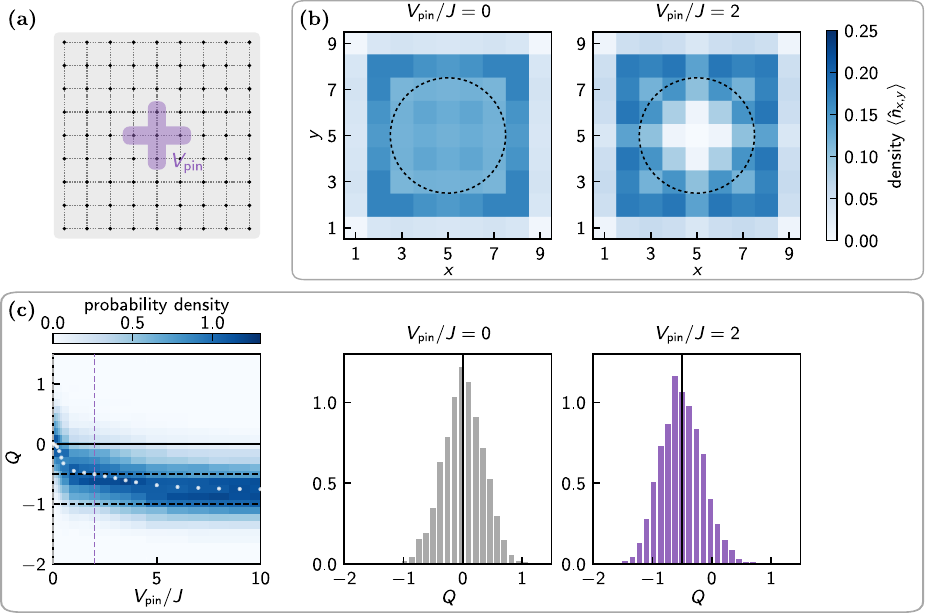}
	\caption{
		\emph{Large square}: 
        \textbf{(a)} Pinning a quasi-hole is possible using for example a cross-shaped pinning potential of strength $V_{\rm pin}$ at the center of a system of four coupled $4\times 4$-patches, resulting in density patterns visualized in \textbf{(b)}.
		\textbf{(c)} Taking $N_{\rm shots}=2000$ snapshots and evaluating the pinned charge Eq.~\eqref{eq:ChargeOperator}, we can determine the full counting statistics of the charge operator.
        We find that for pinning potentials $\nicefrac{V_{\rm pin}}{J}\approx 2$ a fractional charge $q_{\rm qh}=-\nicefrac{1}{2}$ is pinned, whereas weaker (stronger) pinning potentials result in no (more than one) quasi-hole being pinned.
        While the mean value of the charge operator (white dots) is not sufficient to conclusively reveal this distinction, the full counting statistics provides clear evidence.
        On the right, we exemplify this behavior for two values of the pinning potential indicated by the correspondingly colored vertical lines in the left panel.
		Data is given for $\nicefrac{U}{J} = \infty$ and $\nicefrac{\xi}{a}=2.5$ (visualized by the dotted circles in \textbf{(b)}).
	}
	\label{fig:4PatchesSquare_Quasihole}
\end{figure*}

We start our analysis by performing DMRG ground state searches for the Hamiltonian $\hat{\mathcal{H}}_{\rm square}(V_{\rm barrier}=0)$ introduced above with an additional cross-shaped potential of strength $V_{\rm pin}$ added on the five center sites, see Fig.~\ref{fig:4PatchesSquare_Quasihole}\textbf{(a)}.
Such a pinning potential exploits the patchwork structure proposed before and seems natural when trying to quasi-adiabatically grow a quasi-hole state in large squares.

We evaluate the ground state density $\braket{\hat{n}_{x,y}}$ for different strengths of the pinning potential $\nicefrac{V_{\rm pin}}{J}$ and define the charge operator~\cite{Kivelson1982,Bibo2020}
\begin{equation}
	\hat{Q} = \sum_{x,y} \mathrm{e}^{-(\vec{r}-\vec{r}_0)^2/\xi^2} \left(\hat{n}_{\vec{r}} - n^{(0)}_{\vec{r}}\right),
	\label{eq:ChargeOperator}
\end{equation}
where $\vec{r}_0=(5,5)$ denotes the central site of the pinning potential and $n^{(0)}$ is the density for $V_{\rm pin}=0$.
Furthermore, we introduced the width $\nicefrac{\xi}{a}=2.5$ of the Gaussian envelope function, which has to be larger than the correlation length ($\xi_{\rm corr} \approx \ell_B$ for the Laughlin state) and significantly smaller than the system size~\cite{Kivelson1982,Bibo2020}.
Our choice of $\xi$ is such that it clearly encircles most of the density depletion but still remains sufficiently far away from the edge of the system, see Fig.~\ref{fig:4PatchesSquare_Quasihole}\textbf{(b)}.

Using Fock-basis snapshots of the system obtained via a perfect sampling scheme for MPS~\cite{Ferris2012}, we investigate the possibility to observe the fractional charge in quantum gas microscopy experiments.
To this end, we generate $N_{\rm shots}=2000$ snapshots and calculate the pinned charge $\hat{Q}$ for each snapshot separately.
In this way, we obtain the full counting statistics for the charge, which provides additional information beyond the average $\braket{\hat{Q}}$ that is required to confirm the quantization of the quasi-hole charge.

We find that for a pinning potential $\nicefrac{V_{\rm pin}}{J} \approx 2$ the distribution of the pinned charge is peaked around $q_{\rm qh} \approx -\nicefrac{1}{2}$, see Fig.~\ref{fig:4PatchesSquare_Quasihole}\textbf{(c)}, consistent with the pinning of a quasi-hole, where we attribute the deviations from the exact value $-\nicefrac{1}{2}$ to finite-size effects.
In contrast, we find that while the distribution is clearly peaked around $q_{\rm qh} = 0$ for weak pinning potentials, indicating the incompressibility of the Laughlin state, sufficiently strong pinning potentials, $\nicefrac{V_{\rm pin}}{J} \gtrsim 5$, lead to more than one quasi-hole being pinned, or possibly higer-Landau level effects.

We remark that our approach directly probes the fractionalization of the quasi-hole charge, while earlier studies relied on a plateau of the mean pinned charge (indicated by white dots in Fig.~\ref{fig:4PatchesSquare_Quasihole}\textbf{(b)}) for a finite range of pinning strengths~\cite{Macaluso2020,Raciunas2018,Wang2022}.
We find that finite-size effects tend to destroy this plateau, while the full counting statistics still allows for a conclusive extraction of the quasi-hole charge.
While strictly speaking it is necessary to confirm that the peak in the full counting statistics becomes sharper as $L_x, L_y \to \infty$ to unambiguously prove the fractionalized nature of the pinned charge, already relatively small systems like those studied here - and, more importantly, available to near-term quantum simulators - show strong signatures of the fractional quasi-hole charge.
Accordingly, we conclude that the direct observation of a quantized fractional charge is now within reach.

\subsubsection{Time-evolution}
Having established the ground state in the presence of the cross-shaped pinning potential $\nicefrac{V_{\rm pin}}{J} = 2$ as a quasi-hole state of the Laughlin state, we now turn to its quasi-adiabatic preparation.
As before, we start from a product state of four $2$-particle Laughlin states and ramp down the potential barrier on almost all sites except those forming the pinning potential.
In particular, we fix the strength of the pinning potential to $\nicefrac{V_{\rm pin}}{J}=2$ independently of the barrier height $\nicefrac{V_{\rm barrier}(t)}{J}$.

\begin{figure*}
	\centering
	\includegraphics{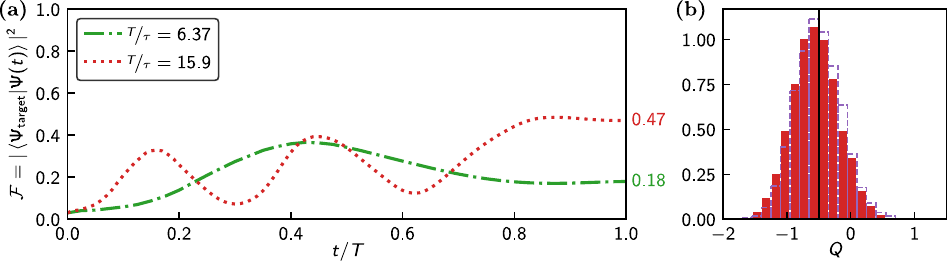}
	\caption{
		\emph{Large square}:
		\textbf{(a)} Target-state fidelity $\mathcal{F}$ for a linearly decreasing barrier height in the presence of a cross-shaped repulsive pinning potential of strength $\nicefrac{V_{\rm pin}}{J} = 2$ in the center.
		We find significant overlap of the time-evolved state with the quasi-hole target-state for both preparation times.
		\textbf{(b)} The full counting statistics of the pinned charge $Q$ obtained from $N_{\rm shots}=2000$ snapshots clearly shows the quantization of the quasi-hole charge to $q_{\rm qh}=-\nicefrac{1}{2}$ both for the time evolved state (filled red bars, $\nicefrac{T}{\tau}=15.9$) and the target-state obtained via DMRG (purple dashed line).
		Data is given for $\nicefrac{U}{J} = \infty$ and $\nicefrac{\xi}{a} = 2.5$.
	}
	\label{fig:4PatchesSquare_QuasiholeOverlap}
\end{figure*}

We again calculate the target-state fidelity along the ramp and find a final fidelity of $\mathcal{F}(T) = 0.18~(0.47)$ for ramp time $\nicefrac{T}{\tau}=6.37~(15.9)$, see Fig.~\ref{fig:4PatchesSquare_QuasiholeOverlap}.
Note in particular, that these fidelities are significantly higher than those obtained for the preparation of the Laughlin state itself, see Fig.~\ref{fig:Supersquare:Overlap}.
This is likely to be a result of the spatial structure of the initial state favoring the quasi-hole state as is also visible from the increased overlap of the initial state $\ket{\Psi(t=0)}$ with the quasi-hole state, see Fig.~\ref{fig:4PatchesSquare_QuasiholeOverlap}, compared to the Laughlin state, see Fig.~\ref{fig:Supersquare:Overlap}.

To further confirm the successful preparation of the quasi-hole state, we again determine the full counting statistics of the charge $\hat{Q}$ defined in Eq.~\eqref{eq:ChargeOperator} for the final state of our preparation protocol.
We find the distribution to be clearly peaked around $q_{\rm qh} = -\nicefrac{1}{2}$ and to agree well with the one extracted from the target-state obtained in our earlier DMRG simulations.
Therefore, we conclude that the reduced fidelity $\mathcal{F}(T) < 1$ of the time-evolved state does not affect the quantization of the quasi-hole charge significantly, so that it is a promising candidate to identify the quasi-hole in cold atom experiments.

In summary, we find that it is possible to grow both a Laughlin state of $N=8$ atoms as well as a quasi-hole state with fractional charge $q_{\rm qh}=-\nicefrac{1}{2}$ using a simple preparation protocol starting from four $4\times4$-patches.

\section{Benchmark: Coupling Chains}
To benchmark the efficiency of our patchwork growing schemes, we compare the results to a preparation protocol where initially decoupled one-dimensional chains are slowly coupled~\cite{He2017}.
To this end, we study anisotropically coupled systems described by the Hamiltonian
\begin{widetext}
	\begin{equation}
		\begin{aligned}
			\hat{\mathcal{H}} = &- \sum_{x,y} \left(J_x \hat{a}^{\dagger}_{x+1,y}\hat{a}_{x,y}^{\vphantom\dagger} + J_y \mathrm{e}^{2\pi i \alpha x}\hat{a}^{\dagger}_{x,y+1}\hat{a}_{x,y}^{\vphantom\dagger} + \mathrm{H.c.}\right) +\frac{U}{2}\sum_{x,y} \hat{n}_{x,y}\left(\hat{n}_{x,y}-1\right),
		\end{aligned}
	\end{equation}
\end{widetext}
which resembles the Hamiltonian studied above for \mbox{$J_y = J_x = J$}.

\begin{figure}
	\centering
	\includegraphics{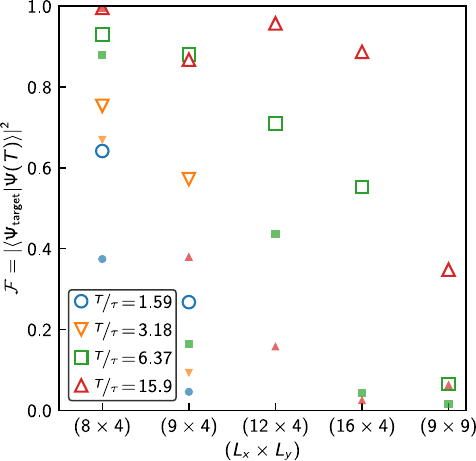}
	\caption{
		Comparison of our proposed patchwork growing schemes (empty symbols) to a scheme coupling chains (faded filled symbols) for different ramp times $T$ and system sizes $(L_x\times L_y)$ discussed above, where we always show the best preparation protocol identified before.
		For larger systems our patchwork approach yields higher target-state fidelities at the end of the ramp.
		In particular, the advantage of this approach becomes clearly visible for longer systems, $L_x \gg L_y$, which clearly shows the scalability of our protocol.
		All data is given for $\nicefrac{U}{J}=\infty$.
	}
	\label{fig:Comparison_Chains}
\end{figure}

We determine the ground state of the system in the completely decoupled limit, $\nicefrac{J_y}{J_x}=0$.
Afterwards, we linearly turn on the hopping in the $y$-direction, i.e.
\begin{equation}
	J_y(t) = \frac{t}{T}~J_x.
\end{equation}
Performing time-evolution simulations using ED (for $(L_x \times L_y) = (8, 4), (9,4)$) and TDVP (for $(L_x \times L_y) = (12,4), (16,4), (9,9)$) as before, we calculate the target-state fidelity for different ramp times, see Fig.~\ref{fig:Comparison_Chains}.
Using the same parameters as before, we compare these results to our earlier findings and conclude that our patchwork approach is clearly favorable.
In particular, for elongated systems, $L_x \gg L_y$, we find a significant advantage of the patchwork approach, clearly showing the scalability of our protocol.

We note that the qualitatively different initial states make a one-to-one comparison of the final fidelities challenging.
Experimental advantages in the preparation of one-dimensional superfluids in decoupled chains makes the simple preparation protocol used as benchmark here appealing, despite the substantially lower fidelities in a perfect numerical environment.

\section{Conclusion and Outlook}
Performing static and time-evolution simulations, we found evidence that various different preparation protocols are well-suited to prepare extended Laughlin states by connecting together copies of the recently realized two-particle state on $4\times4$ sites.
In particular, time-evolution simulations found target-state fidelities above $80\%$ for connecting two patches for preparation times $\nicefrac{T}{\tau} \geq 6$.
Extended chains of up to $20\times4$ sites could be grown with similar fidelities by simultaneously turning on couplings between neighboring patches over a time $\nicefrac{T}{\tau}=15.9$.
Alternative protocols investigated here did not yield higher fidelity, but we believe further optimization of the ramp to be worthwhile in the future.

Finally, we successfully investigated the possibility to couple four patches in a square geometry by linearly turning off an initial potential barrier and prepared a Laughlin state as well as a quasi-hole state.
In particular, we find a fractionally charged quasi-hole around the applied pinning potential both in ground state simulations as well as the time-evolved state.
We emphasize the robustness of the quantized charge despite a target-state fidelity of $47\%$, so that it is experimentally realistic to employ our preparation protocol in the preparation and detection of emergent anyons.
We compared our patchwork approach to an approach of coupling chains and found that the patchwork approach in general gives higher target-state fidelities.

While our results demonstrate the scalability of a patchwork approach, further questions remain.
In particular, it is always desirable to reach even higher fidelities at shorter ramp times to reduce heating and avoid the loss of coherence.
Especially in the case of many patches we believe an optimization of the different couplings to be profitable in the future.
Here, state-of-the-art machine learning techniques might be useful to find particularly efficient paths in parameter space~\cite{Wigley2016,Vendeiro2022,Xie2022,Blatz2023}.
Such an approach might even take into account experimental details which are difficult to investigate on an abstract level.

In light of potential applications, new opportunities unique to cold atom quantum simulators become  available as larger systems are now within reach.
For example, growing long chains will allow for direct measurements of the central charge, thus providing direct evidence of the topological nature of the state.
Furthermore, access to large systems in a square geometry might give the opportunity to directly probe braiding properties of quasi-particles and quasi-holes of the Laughlin state.
In conclusion, building on the latest milestone of realizing a two-atom Laughlin state, exciting new directions are just opening up based on patchwork state preparation approaches.

\begin{acknowledgments}
	The authors would like to thank T.~Blatz, A.~Bohrdt, J.~Dalibard, J.~L\'eonard, and B.~Wang for fruitful discussions.
	FP and FG acknowledge funding by the Deutsche Forschungsgemeinschaft (DFG, German Research Foundation) under Germany's Excellence Strategy -- EXC-2111 -- 390814868, and via Research Unit FOR 2414 under project number 277974659, and from the European Research Council (ERC) under the European Union's Horizon 2020 research and innovation programm (Grant Agreement no 948141) -- ERC Starting Grant SimUcQuam.
    NG acknowledges support by the FRS-FNRS (Belgium), the ERC Grant LATIS and the EOS project CHEQS.
    JK, BBH, and MG acknowledge funding by the ONR grant number N000114-18-1-2863.
\end{acknowledgments}

\appendix
\section{Scaling of the Many-Body Gap in Super-Chains\label{Appendix:ScalingGap}}

\begin{figure*}
	\centering
	\includegraphics{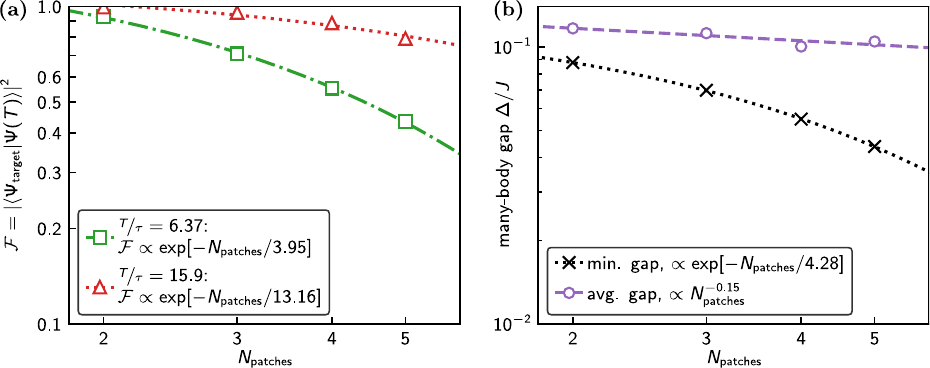}
	\caption{
		\textbf{(a)} Target-state fidelity at the end of the ramp as a function of the number of patches $N_{\rm patches}$ for a scheme linearly coupling all patches simultaneously as discussed in the main text.
		We find that the fidelity decays exponentially as more patches are added, where the decay is significantly suppressed for sufficiently long preparation times.
		\textbf{(b)} The minimal many-body gap along the preparation path also shows an exponential decay as function of $N_{\rm patches}$, in agreement with the existence of gapless edge modes in the thermodynamic limit.
		In contrast, the average gap along the preparation path only decays with algebraically with a small exponent, hence resulting in a favorable scaling of the target-state fidelity for a fixed preparation time.
		Data is given for $\nicefrac{U}{J} = \infty$ and $\alpha=0.3$ in all cases.
	}
	\label{fig:ScalingMBGap}
\end{figure*}

As discussed in the main text, for fixed preparation times the target-state fidelity in extended super-chains decreases with an increasing number of patches $N_{\rm patches}$, see Fig.~\ref{fig:ScalingMBGap}\textbf{(a)}.
While from our simulations we conclude that the overlaps decay exponentially with the number of patches, the data indicates a sufficiently slow decay especially for long preparation times.

Inspired by the adiabatic theorem, the question arises how this relates to the many-body gap along the preparation path.
For fixed flux per plaquette $\alpha=0.3$ the minimal value of the many-body gap $\Delta$ decreases as more patches are added (see Fig.~\ref{fig:ScalingMBGap}\textbf{(b)}), as is expected for the ultimately gapless edge mode in an infinite system.
In particular, we find an exponential decay of the minimal gap as a function of the number of patches $N_{\rm patches}$, which clearly indicates the exponential scaling of the finite-size edge gap.

In contrast, taking into account the entire preparation path it seems meaningful to average the gap along the linear ramps discussed in the main text, i.e. \mbox{$\Delta_{\rm avg} = \int_{0}^{J} \Delta(J_{\rm coupling})~\mathrm{d}J_{\rm coupling}$}.
For the averaged gap we find a slow algebraic decay as number of patches.
As ultimately this averaged gap is expected to limit the ramp speed according to the adiabatic theorem, it seems realistic to grow extended super-chains, especially when further optimizing the ramp speed according to the instantaneous many-body gap.

%
%

\begin{thebibliography}{58}%
	\makeatletter
	\providecommand \@ifxundefined [1]{%
		\@ifx{#1\undefined}
	}%
	\providecommand \@ifnum [1]{%
		\ifnum #1\expandafter \@firstoftwo
		\else \expandafter \@secondoftwo
		\fi
	}%
	\providecommand \@ifx [1]{%
		\ifx #1\expandafter \@firstoftwo
		\else \expandafter \@secondoftwo
		\fi
	}%
	\providecommand \natexlab [1]{#1}%
	\providecommand \enquote  [1]{``#1''}%
	\providecommand \bibnamefont  [1]{#1}%
	\providecommand \bibfnamefont [1]{#1}%
	\providecommand \citenamefont [1]{#1}%
	\providecommand \href@noop [0]{\@secondoftwo}%
	\providecommand \href [0]{\begingroup \@sanitize@url \@href}%
	\providecommand \@href[1]{\@@startlink{#1}\@@href}%
	\providecommand \@@href[1]{\endgroup#1\@@endlink}%
	\providecommand \@sanitize@url [0]{\catcode `\\12\catcode `\$12\catcode
		`\&12\catcode `\#12\catcode `\^12\catcode `\_12\catcode `\%12\relax}%
	\providecommand \@@startlink[1]{}%
	\providecommand \@@endlink[0]{}%
	\providecommand \url  [0]{\begingroup\@sanitize@url \@url }%
	\providecommand \@url [1]{\endgroup\@href {#1}{\urlprefix }}%
	\providecommand \urlprefix  [0]{URL }%
	\providecommand \Eprint [0]{\href }%
	\providecommand \doibase [0]{https://doi.org/}%
	\providecommand \selectlanguage [0]{\@gobble}%
	\providecommand \bibinfo  [0]{\@secondoftwo}%
	\providecommand \bibfield  [0]{\@secondoftwo}%
	\providecommand \translation [1]{[#1]}%
	\providecommand \BibitemOpen [0]{}%
	\providecommand \bibitemStop [0]{}%
	\providecommand \bibitemNoStop [0]{.\EOS\space}%
	\providecommand \EOS [0]{\spacefactor3000\relax}%
	\providecommand \BibitemShut  [1]{\csname bibitem#1\endcsname}%
	\let\auto@bib@innerbib\@empty
	\bibitem [{\citenamefont {Tsui}\ \emph {et~al.}(1982)\citenamefont {Tsui},
		\citenamefont {Stormer},\ and\ \citenamefont {Gossard}}]{Tsui1982}%
	\BibitemOpen
	\bibfield  {author} {\bibinfo {author} {\bibfnamefont {D.~C.}\ \bibnamefont
			{Tsui}}, \bibinfo {author} {\bibfnamefont {H.~L.}\ \bibnamefont {Stormer}},\
		and\ \bibinfo {author} {\bibfnamefont {A.~C.}\ \bibnamefont {Gossard}},\
	}\bibfield  {title} {\bibinfo {title} {{Two-Dimensional Magnetotransport in
				the Extreme Quantum Limit}},\ }\href
	{https://doi.org/10.1103/physrevlett.48.1559} {\bibfield  {journal} {\bibinfo
			{journal} {Physical Review Letters}\ }\textbf {\bibinfo {volume} {48}},\
		\bibinfo {pages} {1559} (\bibinfo {year} {1982})}\BibitemShut {NoStop}%
	\bibitem [{\citenamefont {Laughlin}(1983)}]{Laughlin1983}%
	\BibitemOpen
	\bibfield  {author} {\bibinfo {author} {\bibfnamefont {R.~B.}\ \bibnamefont
			{Laughlin}},\ }\bibfield  {title} {\bibinfo {title} {{Anomalous Quantum Hall
				Effect: An Incompressible Quantum Fluid with Fractionally Charged
				Excitations}},\ }\href {https://doi.org/10.1103/physrevlett.50.1395}
	{\bibfield  {journal} {\bibinfo  {journal} {Physical Review Letters}\
		}\textbf {\bibinfo {volume} {50}},\ \bibinfo {pages} {1395} (\bibinfo {year}
		{1983})}\BibitemShut {NoStop}%
	\bibitem [{\citenamefont {Halperin}(1984)}]{Halperin1984}%
	\BibitemOpen
	\bibfield  {author} {\bibinfo {author} {\bibfnamefont {B.~I.}\ \bibnamefont
			{Halperin}},\ }\bibfield  {title} {\bibinfo {title} {{Statistics of
				Quasiparticles and the Hierarchy of Fractional Quantized Hall States}},\
	}\href {https://doi.org/10.1103/physrevlett.52.1583} {\bibfield  {journal}
		{\bibinfo  {journal} {Physical Review Letters}\ }\textbf {\bibinfo {volume}
			{52}},\ \bibinfo {pages} {1583} (\bibinfo {year} {1984})}\BibitemShut
	{NoStop}%
	\bibitem [{\citenamefont {Arovas}\ \emph {et~al.}(1984)\citenamefont {Arovas},
		\citenamefont {Schrieffer},\ and\ \citenamefont {Wilczek}}]{Arovas1984}%
	\BibitemOpen
	\bibfield  {author} {\bibinfo {author} {\bibfnamefont {D.}~\bibnamefont
			{Arovas}}, \bibinfo {author} {\bibfnamefont {J.~R.}\ \bibnamefont
			{Schrieffer}},\ and\ \bibinfo {author} {\bibfnamefont {F.}~\bibnamefont
			{Wilczek}},\ }\bibfield  {title} {\bibinfo {title} {{Fractional Statistics
				and the Quantum Hall Effect}},\ }\href
	{https://doi.org/10.1103/physrevlett.53.722} {\bibfield  {journal} {\bibinfo
			{journal} {Physical Review Letters}\ }\textbf {\bibinfo {volume} {53}},\
		\bibinfo {pages} {722} (\bibinfo {year} {1984})}\BibitemShut {NoStop}%
	\bibitem [{\citenamefont {Aidelsburger}\ \emph {et~al.}(2013)\citenamefont
		{Aidelsburger}, \citenamefont {Atala}, \citenamefont {Lohse}, \citenamefont
		{Barreiro}, \citenamefont {Paredes},\ and\ \citenamefont
		{Bloch}}]{Aidelsburger2013}%
	\BibitemOpen
	\bibfield  {author} {\bibinfo {author} {\bibfnamefont {M.}~\bibnamefont
			{Aidelsburger}}, \bibinfo {author} {\bibfnamefont {M.}~\bibnamefont {Atala}},
		\bibinfo {author} {\bibfnamefont {M.}~\bibnamefont {Lohse}}, \bibinfo
		{author} {\bibfnamefont {J.~T.}\ \bibnamefont {Barreiro}}, \bibinfo {author}
		{\bibfnamefont {B.}~\bibnamefont {Paredes}},\ and\ \bibinfo {author}
		{\bibfnamefont {I.}~\bibnamefont {Bloch}},\ }\bibfield  {title} {\bibinfo
		{title} {{Realization of the Hofstadter Hamiltonian with Ultracold Atoms in
				Optical Lattices}},\ }\href {https://doi.org/10.1103/physrevlett.111.185301}
	{\bibfield  {journal} {\bibinfo  {journal} {Physical Review Letters}\
		}\textbf {\bibinfo {volume} {111}},\ \bibinfo {pages} {185301} (\bibinfo
		{year} {2013})}\BibitemShut {NoStop}%
	\bibitem [{\citenamefont {Miyake}\ \emph {et~al.}(2013)\citenamefont {Miyake},
		\citenamefont {Siviloglou}, \citenamefont {Kennedy}, \citenamefont {Burton},\
		and\ \citenamefont {Ketterle}}]{Miyake2013}%
	\BibitemOpen
	\bibfield  {author} {\bibinfo {author} {\bibfnamefont {H.}~\bibnamefont
			{Miyake}}, \bibinfo {author} {\bibfnamefont {G.~A.}\ \bibnamefont
			{Siviloglou}}, \bibinfo {author} {\bibfnamefont {C.~J.}\ \bibnamefont
			{Kennedy}}, \bibinfo {author} {\bibfnamefont {W.~C.}\ \bibnamefont
			{Burton}},\ and\ \bibinfo {author} {\bibfnamefont {W.}~\bibnamefont
			{Ketterle}},\ }\bibfield  {title} {\bibinfo {title} {{Realizing the Harper
				Hamiltonian with Laser-Assisted Tunneling in Optical Lattices}},\ }\href
	{https://doi.org/10.1103/physrevlett.111.185302} {\bibfield  {journal}
		{\bibinfo  {journal} {Physical Review Letters}\ }\textbf {\bibinfo {volume}
			{111}},\ \bibinfo {pages} {185302} (\bibinfo {year} {2013})}\BibitemShut
	{NoStop}%
	\bibitem [{\citenamefont {Tai}\ \emph {et~al.}(2017)\citenamefont {Tai},
		\citenamefont {Lukin}, \citenamefont {Rispoli}, \citenamefont {Schittko},
		\citenamefont {Menke}, \citenamefont {Borgnia}, \citenamefont {Preiss},
		\citenamefont {Grusdt}, \citenamefont {Kaufman},\ and\ \citenamefont
		{Greiner}}]{Tai2017}%
	\BibitemOpen
	\bibfield  {author} {\bibinfo {author} {\bibfnamefont {M.~E.}\ \bibnamefont
			{Tai}}, \bibinfo {author} {\bibfnamefont {A.}~\bibnamefont {Lukin}}, \bibinfo
		{author} {\bibfnamefont {M.}~\bibnamefont {Rispoli}}, \bibinfo {author}
		{\bibfnamefont {R.}~\bibnamefont {Schittko}}, \bibinfo {author}
		{\bibfnamefont {T.}~\bibnamefont {Menke}}, \bibinfo {author} {\bibfnamefont
			{D.}~\bibnamefont {Borgnia}}, \bibinfo {author} {\bibfnamefont {P.~M.}\
			\bibnamefont {Preiss}}, \bibinfo {author} {\bibfnamefont {F.}~\bibnamefont
			{Grusdt}}, \bibinfo {author} {\bibfnamefont {A.~M.}\ \bibnamefont
			{Kaufman}},\ and\ \bibinfo {author} {\bibfnamefont {M.}~\bibnamefont
			{Greiner}},\ }\bibfield  {title} {\bibinfo {title} {{Microscopy of the
				interacting Harper{\textendash}Hofstadter model in the two-body limit}},\
	}\href {https://doi.org/10.1038/nature22811} {\bibfield  {journal} {\bibinfo
			{journal} {Nature}\ }\textbf {\bibinfo {volume} {546}},\ \bibinfo {pages}
		{519} (\bibinfo {year} {2017})}\BibitemShut {NoStop}%
	\bibitem [{\citenamefont {Jotzu}\ \emph {et~al.}(2014)\citenamefont {Jotzu},
		\citenamefont {Messer}, \citenamefont {Desbuquois}, \citenamefont {Lebrat},
		\citenamefont {Uehlinger}, \citenamefont {Greif},\ and\ \citenamefont
		{Esslinger}}]{Jotzu2014}%
	\BibitemOpen
	\bibfield  {author} {\bibinfo {author} {\bibfnamefont {G.}~\bibnamefont
			{Jotzu}}, \bibinfo {author} {\bibfnamefont {M.}~\bibnamefont {Messer}},
		\bibinfo {author} {\bibfnamefont {R.}~\bibnamefont {Desbuquois}}, \bibinfo
		{author} {\bibfnamefont {M.}~\bibnamefont {Lebrat}}, \bibinfo {author}
		{\bibfnamefont {T.}~\bibnamefont {Uehlinger}}, \bibinfo {author}
		{\bibfnamefont {D.}~\bibnamefont {Greif}},\ and\ \bibinfo {author}
		{\bibfnamefont {T.}~\bibnamefont {Esslinger}},\ }\bibfield  {title} {\bibinfo
		{title} {{Experimental realization of the topological Haldane model with
				ultracold fermions}},\ }\href {https://doi.org/10.1038/nature13915}
	{\bibfield  {journal} {\bibinfo  {journal} {Nature}\ }\textbf {\bibinfo
			{volume} {515}},\ \bibinfo {pages} {237} (\bibinfo {year}
		{2014})}\BibitemShut {NoStop}%
	\bibitem [{\citenamefont {Fl\"aschner}\ \emph {et~al.}(2016)\citenamefont
		{Fl\"aschner}, \citenamefont {Rem}, \citenamefont {Tarnowski}, \citenamefont
		{Vogel}, \citenamefont {Luhmann}, \citenamefont {Sengstock},\ and\
		\citenamefont {Weitenberg}}]{Flaeschner2016}%
	\BibitemOpen
	\bibfield  {author} {\bibinfo {author} {\bibfnamefont {N.}~\bibnamefont
			{Fl\"aschner}}, \bibinfo {author} {\bibfnamefont {B.~S.}\ \bibnamefont
			{Rem}}, \bibinfo {author} {\bibfnamefont {M.}~\bibnamefont {Tarnowski}},
		\bibinfo {author} {\bibfnamefont {D.}~\bibnamefont {Vogel}}, \bibinfo
		{author} {\bibfnamefont {D.-S.}\ \bibnamefont {Luhmann}}, \bibinfo {author}
		{\bibfnamefont {K.}~\bibnamefont {Sengstock}},\ and\ \bibinfo {author}
		{\bibfnamefont {C.}~\bibnamefont {Weitenberg}},\ }\bibfield  {title}
	{\bibinfo {title} {{Experimental reconstruction of the Berry curvature in a
				Floquet Bloch band}},\ }\href {https://doi.org/10.1126/science.aad4568}
	{\bibfield  {journal} {\bibinfo  {journal} {Science}\ }\textbf {\bibinfo
			{volume} {352}},\ \bibinfo {pages} {1091} (\bibinfo {year}
		{2016})}\BibitemShut {NoStop}%
	\bibitem [{\citenamefont {Goldman}\ \emph {et~al.}(2016)\citenamefont
		{Goldman}, \citenamefont {Budich},\ and\ \citenamefont
		{Zoller}}]{Goldman2016}%
	\BibitemOpen
	\bibfield  {author} {\bibinfo {author} {\bibfnamefont {N.}~\bibnamefont
			{Goldman}}, \bibinfo {author} {\bibfnamefont {J.~C.}\ \bibnamefont
			{Budich}},\ and\ \bibinfo {author} {\bibfnamefont {P.}~\bibnamefont
			{Zoller}},\ }\bibfield  {title} {\bibinfo {title} {Topological quantum matter
			with ultracold gases in optical lattices},\ }\href
	{https://doi.org/10.1038/nphys3803} {\bibfield  {journal} {\bibinfo
			{journal} {Nature Physics}\ }\textbf {\bibinfo {volume} {12}},\ \bibinfo
		{pages} {639} (\bibinfo {year} {2016})}\BibitemShut {NoStop}%
	\bibitem [{\citenamefont {Aidelsburger}\ \emph {et~al.}(2018)\citenamefont
		{Aidelsburger}, \citenamefont {Nascimbene},\ and\ \citenamefont
		{Goldman}}]{Aidelsburger2018}%
	\BibitemOpen
	\bibfield  {author} {\bibinfo {author} {\bibfnamefont {M.}~\bibnamefont
			{Aidelsburger}}, \bibinfo {author} {\bibfnamefont {S.}~\bibnamefont
			{Nascimbene}},\ and\ \bibinfo {author} {\bibfnamefont {N.}~\bibnamefont
			{Goldman}},\ }\bibfield  {title} {\bibinfo {title} {Artificial gauge fields
			in materials and engineered systems},\ }\href
	{https://doi.org/10.1016/j.crhy.2018.03.002} {\bibfield  {journal} {\bibinfo
			{journal} {Comptes Rendus Physique}\ }\textbf {\bibinfo {volume} {19}},\
		\bibinfo {pages} {394} (\bibinfo {year} {2018})}\BibitemShut {NoStop}%
	\bibitem [{\citenamefont {Cooper}\ \emph {et~al.}(2019)\citenamefont {Cooper},
		\citenamefont {Dalibard},\ and\ \citenamefont {Spielman}}]{Cooper2019}%
	\BibitemOpen
	\bibfield  {author} {\bibinfo {author} {\bibfnamefont {N.~R.}\ \bibnamefont
			{Cooper}}, \bibinfo {author} {\bibfnamefont {J.}~\bibnamefont {Dalibard}},\
		and\ \bibinfo {author} {\bibfnamefont {I.~B.}\ \bibnamefont {Spielman}},\
	}\bibfield  {title} {\bibinfo {title} {Topological bands for ultracold
			atoms},\ }\href {https://doi.org/10.1103/revmodphys.91.015005} {\bibfield
		{journal} {\bibinfo  {journal} {Reviews of Modern Physics}\ }\textbf
		{\bibinfo {volume} {91}},\ \bibinfo {pages} {015005} (\bibinfo {year}
		{2019})}\BibitemShut {NoStop}%
	\bibitem [{\citenamefont {Clark}\ \emph {et~al.}(2020)\citenamefont {Clark},
		\citenamefont {Schine}, \citenamefont {Baum}, \citenamefont {Jia},\ and\
		\citenamefont {Simon}}]{Clark2020}%
	\BibitemOpen
	\bibfield  {author} {\bibinfo {author} {\bibfnamefont {L.~W.}\ \bibnamefont
			{Clark}}, \bibinfo {author} {\bibfnamefont {N.}~\bibnamefont {Schine}},
		\bibinfo {author} {\bibfnamefont {C.}~\bibnamefont {Baum}}, \bibinfo {author}
		{\bibfnamefont {N.}~\bibnamefont {Jia}},\ and\ \bibinfo {author}
		{\bibfnamefont {J.}~\bibnamefont {Simon}},\ }\bibfield  {title} {\bibinfo
		{title} {{Observation of Laughlin states made of light}},\ }\href
	{https://doi.org/10.1038/s41586-020-2318-5} {\bibfield  {journal} {\bibinfo
			{journal} {Nature}\ }\textbf {\bibinfo {volume} {582}},\ \bibinfo {pages}
		{41} (\bibinfo {year} {2020})}\BibitemShut {NoStop}%
	\bibitem [{\citenamefont {L{\'{e}}onard}\ \emph {et~al.}(2023)\citenamefont
		{L{\'{e}}onard}, \citenamefont {Kim}, \citenamefont {Kwan}, \citenamefont
		{Segura}, \citenamefont {Grusdt}, \citenamefont {Repellin}, \citenamefont
		{Goldman},\ and\ \citenamefont {Greiner}}]{Leonard2023a}%
	\BibitemOpen
	\bibfield  {author} {\bibinfo {author} {\bibfnamefont {J.}~\bibnamefont
			{L{\'{e}}onard}}, \bibinfo {author} {\bibfnamefont {S.}~\bibnamefont {Kim}},
		\bibinfo {author} {\bibfnamefont {J.}~\bibnamefont {Kwan}}, \bibinfo {author}
		{\bibfnamefont {P.}~\bibnamefont {Segura}}, \bibinfo {author} {\bibfnamefont
			{F.}~\bibnamefont {Grusdt}}, \bibinfo {author} {\bibfnamefont
			{C.}~\bibnamefont {Repellin}}, \bibinfo {author} {\bibfnamefont
			{N.}~\bibnamefont {Goldman}},\ and\ \bibinfo {author} {\bibfnamefont
			{M.}~\bibnamefont {Greiner}},\ }\bibfield  {title} {\bibinfo {title}
		{{Realization of a fractional quantum Hall state with ultracold atoms}},\
	}\href {https://doi.org/10.1038/s41586-023-06122-4} {\bibfield  {journal}
		{\bibinfo  {journal} {Nature}\ }\textbf {\bibinfo {volume} {619}},\ \bibinfo
		{pages} {495} (\bibinfo {year} {2023})}\BibitemShut {NoStop}%
	\bibitem [{\citenamefont {Popp}\ \emph {et~al.}(2004)\citenamefont {Popp},
		\citenamefont {Paredes},\ and\ \citenamefont {Cirac}}]{Popp2004}%
	\BibitemOpen
	\bibfield  {author} {\bibinfo {author} {\bibfnamefont {M.}~\bibnamefont
			{Popp}}, \bibinfo {author} {\bibfnamefont {B.}~\bibnamefont {Paredes}},\ and\
		\bibinfo {author} {\bibfnamefont {J.~I.}\ \bibnamefont {Cirac}},\ }\bibfield
	{title} {\bibinfo {title} {{Adiabatic path to fractional quantum Hall states
				of a few bosonic atoms}},\ }\href
	{https://doi.org/10.1103/physreva.70.053612} {\bibfield  {journal} {\bibinfo
			{journal} {Physical Review A}\ }\textbf {\bibinfo {volume} {70}},\ \bibinfo
		{pages} {053612} (\bibinfo {year} {2004})}\BibitemShut {NoStop}%
	\bibitem [{\citenamefont {Barkeshli}\ and\ \citenamefont
		{McGreevy}(2014)}]{Barkeshli2014}%
	\BibitemOpen
	\bibfield  {author} {\bibinfo {author} {\bibfnamefont {M.}~\bibnamefont
			{Barkeshli}}\ and\ \bibinfo {author} {\bibfnamefont {J.}~\bibnamefont
			{McGreevy}},\ }\bibfield  {title} {\bibinfo {title} {{Continuous transition
				between fractional quantum Hall and superfluid states}},\ }\href
	{https://doi.org/10.1103/physrevb.89.235116} {\bibfield  {journal} {\bibinfo
			{journal} {Physical Review B}\ }\textbf {\bibinfo {volume} {89}},\ \bibinfo
		{pages} {235116} (\bibinfo {year} {2014})}\BibitemShut {NoStop}%
	\bibitem [{\citenamefont {Barkeshli}\ \emph {et~al.}(2015)\citenamefont
		{Barkeshli}, \citenamefont {Yao},\ and\ \citenamefont
		{Laumann}}]{Barkeshli2015}%
	\BibitemOpen
	\bibfield  {author} {\bibinfo {author} {\bibfnamefont {M.}~\bibnamefont
			{Barkeshli}}, \bibinfo {author} {\bibfnamefont {N.~Y.}\ \bibnamefont {Yao}},\
		and\ \bibinfo {author} {\bibfnamefont {C.~R.}\ \bibnamefont {Laumann}},\
	}\bibfield  {title} {\bibinfo {title} {{Continuous Preparation of a
				Fractional Chern Insulator}},\ }\href
	{https://doi.org/10.1103/physrevlett.115.026802} {\bibfield  {journal}
		{\bibinfo  {journal} {Physical Review Letters}\ }\textbf {\bibinfo {volume}
			{115}},\ \bibinfo {pages} {026802} (\bibinfo {year} {2015})}\BibitemShut
	{NoStop}%
	\bibitem [{\citenamefont {Motruk}\ and\ \citenamefont
		{Pollmann}(2017)}]{Motruk2017}%
	\BibitemOpen
	\bibfield  {author} {\bibinfo {author} {\bibfnamefont {J.}~\bibnamefont
			{Motruk}}\ and\ \bibinfo {author} {\bibfnamefont {F.}~\bibnamefont
			{Pollmann}},\ }\bibfield  {title} {\bibinfo {title} {{Phase transitions and
				adiabatic preparation of a fractional Chern insulator in a boson cold-atom
				model}},\ }\href {https://doi.org/10.1103/physrevb.96.165107} {\bibfield
		{journal} {\bibinfo  {journal} {Physical Review B}\ }\textbf {\bibinfo
			{volume} {96}},\ \bibinfo {pages} {165107} (\bibinfo {year}
		{2017})}\BibitemShut {NoStop}%
	\bibitem [{\citenamefont {Hudomal}\ \emph {et~al.}(2019)\citenamefont
		{Hudomal}, \citenamefont {Regnault},\ and\ \citenamefont
		{Vasi{\'{c}}}}]{Hudomal2019}%
	\BibitemOpen
	\bibfield  {author} {\bibinfo {author} {\bibfnamefont {A.}~\bibnamefont
			{Hudomal}}, \bibinfo {author} {\bibfnamefont {N.}~\bibnamefont {Regnault}},\
		and\ \bibinfo {author} {\bibfnamefont {I.}~\bibnamefont {Vasi{\'{c}}}},\
	}\bibfield  {title} {\bibinfo {title} {{Bosonic fractional quantum Hall
				states in driven optical lattices}},\ }\href
	{https://doi.org/10.1103/physreva.100.053624} {\bibfield  {journal} {\bibinfo
			{journal} {Physical Review A}\ }\textbf {\bibinfo {volume} {100}},\ \bibinfo
		{pages} {053624} (\bibinfo {year} {2019})}\BibitemShut {NoStop}%
	\bibitem [{\citenamefont {Andrade}\ \emph {et~al.}(2021)\citenamefont
		{Andrade}, \citenamefont {Kasper}, \citenamefont {Lewenstein}, \citenamefont
		{Weitenberg},\ and\ \citenamefont {Gra\ss}}]{Andrade2021}%
	\BibitemOpen
	\bibfield  {author} {\bibinfo {author} {\bibfnamefont {B.}~\bibnamefont
			{Andrade}}, \bibinfo {author} {\bibfnamefont {V.}~\bibnamefont {Kasper}},
		\bibinfo {author} {\bibfnamefont {M.}~\bibnamefont {Lewenstein}}, \bibinfo
		{author} {\bibfnamefont {C.}~\bibnamefont {Weitenberg}},\ and\ \bibinfo
		{author} {\bibfnamefont {T.}~\bibnamefont {Gra\ss}},\ }\bibfield  {title}
	{\bibinfo {title} {{Preparation of the 1/2 Laughlin state with atoms in a
				rotating trap}},\ }\href {https://doi.org/10.1103/physreva.103.063325}
	{\bibfield  {journal} {\bibinfo  {journal} {Physical Review A}\ }\textbf
		{\bibinfo {volume} {103}},\ \bibinfo {pages} {063325} (\bibinfo {year}
		{2021})}\BibitemShut {NoStop}%
	\bibitem [{\citenamefont {He}\ \emph {et~al.}(2017)\citenamefont {He},
		\citenamefont {Grusdt}, \citenamefont {Kaufman}, \citenamefont {Greiner},\
		and\ \citenamefont {Vishwanath}}]{He2017}%
	\BibitemOpen
	\bibfield  {author} {\bibinfo {author} {\bibfnamefont {Y.-C.}\ \bibnamefont
			{He}}, \bibinfo {author} {\bibfnamefont {F.}~\bibnamefont {Grusdt}}, \bibinfo
		{author} {\bibfnamefont {A.}~\bibnamefont {Kaufman}}, \bibinfo {author}
		{\bibfnamefont {M.}~\bibnamefont {Greiner}},\ and\ \bibinfo {author}
		{\bibfnamefont {A.}~\bibnamefont {Vishwanath}},\ }\bibfield  {title}
	{\bibinfo {title} {{Realizing and adiabatically preparing bosonic integer and
				fractional quantum Hall states in optical lattices}},\ }\href
	{https://doi.org/10.1103/physrevb.96.201103} {\bibfield  {journal} {\bibinfo
			{journal} {Physical Review B}\ }\textbf {\bibinfo {volume} {96}},\ \bibinfo
		{pages} {201103(R)} (\bibinfo {year} {2017})}\BibitemShut {NoStop}%
	\bibitem [{\citenamefont {Grusdt}\ \emph {et~al.}(2016)\citenamefont {Grusdt},
		\citenamefont {Yao}, \citenamefont {Abanin}, \citenamefont {Fleischhauer},\
		and\ \citenamefont {Demler}}]{Grusdt2016}%
	\BibitemOpen
	\bibfield  {author} {\bibinfo {author} {\bibfnamefont {F.}~\bibnamefont
			{Grusdt}}, \bibinfo {author} {\bibfnamefont {N.~Y.}\ \bibnamefont {Yao}},
		\bibinfo {author} {\bibfnamefont {D.}~\bibnamefont {Abanin}}, \bibinfo
		{author} {\bibfnamefont {M.}~\bibnamefont {Fleischhauer}},\ and\ \bibinfo
		{author} {\bibfnamefont {E.}~\bibnamefont {Demler}},\ }\bibfield  {title}
	{\bibinfo {title} {Interferometric measurements of many-body topological
			invariants using mobile impurities},\ }\href
	{https://doi.org/10.1038/ncomms11994} {\bibfield  {journal} {\bibinfo
			{journal} {Nature Communications}\ }\textbf {\bibinfo {volume} {7}},\
		\bibinfo {pages} {11994} (\bibinfo {year} {2016})}\BibitemShut {NoStop}%
	\bibitem [{\citenamefont {Nakamura}\ \emph {et~al.}(2020)\citenamefont
		{Nakamura}, \citenamefont {Liang}, \citenamefont {Gardner},\ and\
		\citenamefont {Manfra}}]{Nakamura2020}%
	\BibitemOpen
	\bibfield  {author} {\bibinfo {author} {\bibfnamefont {J.}~\bibnamefont
			{Nakamura}}, \bibinfo {author} {\bibfnamefont {S.}~\bibnamefont {Liang}},
		\bibinfo {author} {\bibfnamefont {G.~C.}\ \bibnamefont {Gardner}},\ and\
		\bibinfo {author} {\bibfnamefont {M.~J.}\ \bibnamefont {Manfra}},\ }\bibfield
	{title} {\bibinfo {title} {Direct observation of anyonic braiding
			statistics},\ }\href {https://doi.org/10.1038/s41567-020-1019-1} {\bibfield
		{journal} {\bibinfo  {journal} {Nature Physics}\ }\textbf {\bibinfo {volume}
			{16}},\ \bibinfo {pages} {931} (\bibinfo {year} {2020})}\BibitemShut
	{NoStop}%
	\bibitem [{\citenamefont {Umucal{\i}lar}(2018)}]{Umucalilar2018}%
	\BibitemOpen
	\bibfield  {author} {\bibinfo {author} {\bibfnamefont {R.~O.}\ \bibnamefont
			{Umucal{\i}lar}},\ }\bibfield  {title} {\bibinfo {title} {Real-space probe
			for lattice quasiholes},\ }\href {https://doi.org/10.1103/physreva.98.063629}
	{\bibfield  {journal} {\bibinfo  {journal} {Physical Review A}\ }\textbf
		{\bibinfo {volume} {98}},\ \bibinfo {pages} {063629} (\bibinfo {year}
		{2018})}\BibitemShut {NoStop}%
	\bibitem [{\citenamefont {Macaluso}\ \emph {et~al.}(2020)\citenamefont
		{Macaluso}, \citenamefont {Comparin}, \citenamefont {Umucal{\i}lar},
		\citenamefont {Gerster}, \citenamefont {Montangero}, \citenamefont {Rizzi},\
		and\ \citenamefont {Carusotto}}]{Macaluso2020}%
	\BibitemOpen
	\bibfield  {author} {\bibinfo {author} {\bibfnamefont {E.}~\bibnamefont
			{Macaluso}}, \bibinfo {author} {\bibfnamefont {T.}~\bibnamefont {Comparin}},
		\bibinfo {author} {\bibfnamefont {R.~O.}\ \bibnamefont {Umucal{\i}lar}},
		\bibinfo {author} {\bibfnamefont {M.}~\bibnamefont {Gerster}}, \bibinfo
		{author} {\bibfnamefont {S.}~\bibnamefont {Montangero}}, \bibinfo {author}
		{\bibfnamefont {M.}~\bibnamefont {Rizzi}},\ and\ \bibinfo {author}
		{\bibfnamefont {I.}~\bibnamefont {Carusotto}},\ }\bibfield  {title} {\bibinfo
		{title} {{Charge and statistics of lattice quasiholes from density
				measurements: A tree tensor network study}},\ }\href
	{https://doi.org/10.1103/physrevresearch.2.013145} {\bibfield  {journal}
		{\bibinfo  {journal} {Physical Review Research}\ }\textbf {\bibinfo {volume}
			{2}},\ \bibinfo {pages} {013145} (\bibinfo {year} {2020})}\BibitemShut
	{NoStop}%
	\bibitem [{\citenamefont {Palm}\ \emph {et~al.}(2022)\citenamefont {Palm},
		\citenamefont {Mardazad}, \citenamefont {Bohrdt}, \citenamefont
		{Schollw\"ock},\ and\ \citenamefont {Grusdt}}]{Palm2022}%
	\BibitemOpen
	\bibfield  {author} {\bibinfo {author} {\bibfnamefont {F.~A.}\ \bibnamefont
			{Palm}}, \bibinfo {author} {\bibfnamefont {S.}~\bibnamefont {Mardazad}},
		\bibinfo {author} {\bibfnamefont {A.}~\bibnamefont {Bohrdt}}, \bibinfo
		{author} {\bibfnamefont {U.}~\bibnamefont {Schollw\"ock}},\ and\ \bibinfo
		{author} {\bibfnamefont {F.}~\bibnamefont {Grusdt}},\ }\bibfield  {title}
	{\bibinfo {title} {{Snapshot-based detection of $\nu=\frac{1}{2}$ Laughlin
				states: Coupled chains and central charge}},\ }\href
	{https://doi.org/10.1103/PhysRevB.106.L081108} {\bibfield  {journal}
		{\bibinfo  {journal} {Physical Review B}\ }\textbf {\bibinfo {volume}
			{106}},\ \bibinfo {pages} {L081108} (\bibinfo {year} {2022})}\BibitemShut
	{NoStop}%
	\bibitem [{\citenamefont {Binanti}\ \emph {et~al.}(2023)\citenamefont
		{Binanti}, \citenamefont {Goldman},\ and\ \citenamefont
		{Repellin}}]{Binanti2023}%
	\BibitemOpen
	\bibfield  {author} {\bibinfo {author} {\bibfnamefont {F.}~\bibnamefont
			{Binanti}}, \bibinfo {author} {\bibfnamefont {N.}~\bibnamefont {Goldman}},\
		and\ \bibinfo {author} {\bibfnamefont {C.}~\bibnamefont {Repellin}},\
	}\href@noop {} {\bibinfo {title} {{Edge mode spectroscopy of fractional Chern
				insulators}}} (\bibinfo {year} {2023}),\ \Eprint
	{https://arxiv.org/abs/2306.01624} {arXiv:2306.01624} \BibitemShut {NoStop}%
	\bibitem [{\citenamefont {Kane}\ \emph {et~al.}(2002)\citenamefont {Kane},
		\citenamefont {Mukhopadhyay},\ and\ \citenamefont {Lubensky}}]{Kane2002}%
	\BibitemOpen
	\bibfield  {author} {\bibinfo {author} {\bibfnamefont {C.~L.}\ \bibnamefont
			{Kane}}, \bibinfo {author} {\bibfnamefont {R.}~\bibnamefont {Mukhopadhyay}},\
		and\ \bibinfo {author} {\bibfnamefont {T.~C.}\ \bibnamefont {Lubensky}},\
	}\bibfield  {title} {\bibinfo {title} {{Fractional Quantum Hall Effect in an
				Array of Quantum Wires}},\ }\href
	{https://doi.org/10.1103/physrevlett.88.036401} {\bibfield  {journal}
		{\bibinfo  {journal} {Physical Review Letters}\ }\textbf {\bibinfo {volume}
			{88}},\ \bibinfo {pages} {036401} (\bibinfo {year} {2002})}\BibitemShut
	{NoStop}%
	\bibitem [{\citenamefont {Teo}\ and\ \citenamefont {Kane}(2014)}]{Teo2014}%
	\BibitemOpen
	\bibfield  {author} {\bibinfo {author} {\bibfnamefont {J.~C.~Y.}\
			\bibnamefont {Teo}}\ and\ \bibinfo {author} {\bibfnamefont {C.~L.}\
			\bibnamefont {Kane}},\ }\bibfield  {title} {\bibinfo {title} {{From Luttinger
				liquid to non-Abelian quantum Hall states}},\ }\href
	{https://doi.org/10.1103/physrevb.89.085101} {\bibfield  {journal} {\bibinfo
			{journal} {Physical Review B}\ }\textbf {\bibinfo {volume} {89}},\ \bibinfo
		{pages} {085101} (\bibinfo {year} {2014})}\BibitemShut {NoStop}%
	\bibitem [{\citenamefont {Grusdt}\ \emph {et~al.}(2014)\citenamefont {Grusdt},
		\citenamefont {Letscher}, \citenamefont {Hafezi},\ and\ \citenamefont
		{Fleischhauer}}]{Grusdt2014a}%
	\BibitemOpen
	\bibfield  {author} {\bibinfo {author} {\bibfnamefont {F.}~\bibnamefont
			{Grusdt}}, \bibinfo {author} {\bibfnamefont {F.}~\bibnamefont {Letscher}},
		\bibinfo {author} {\bibfnamefont {M.}~\bibnamefont {Hafezi}},\ and\ \bibinfo
		{author} {\bibfnamefont {M.}~\bibnamefont {Fleischhauer}},\ }\bibfield
	{title} {\bibinfo {title} {{Topological Growing of Laughlin States in
				Synthetic Gauge Fields}},\ }\href
	{https://doi.org/10.1103/physrevlett.113.155301} {\bibfield  {journal}
		{\bibinfo  {journal} {Physical Review Letters}\ }\textbf {\bibinfo {volume}
			{113}},\ \bibinfo {pages} {155301} (\bibinfo {year} {2014})}\BibitemShut
	{NoStop}%
	\bibitem [{\citenamefont {Homeier}\ \emph {et~al.}(2021)\citenamefont
		{Homeier}, \citenamefont {Schweizer}, \citenamefont {Aidelsburger},
		\citenamefont {Fedorov},\ and\ \citenamefont {Grusdt}}]{Homeier2021}%
	\BibitemOpen
	\bibfield  {author} {\bibinfo {author} {\bibfnamefont {L.}~\bibnamefont
			{Homeier}}, \bibinfo {author} {\bibfnamefont {C.}~\bibnamefont {Schweizer}},
		\bibinfo {author} {\bibfnamefont {M.}~\bibnamefont {Aidelsburger}}, \bibinfo
		{author} {\bibfnamefont {A.}~\bibnamefont {Fedorov}},\ and\ \bibinfo {author}
		{\bibfnamefont {F.}~\bibnamefont {Grusdt}},\ }\bibfield  {title} {\bibinfo
		{title} {{$\mathbb{Z}_2$ lattice gauge theories and Kitaev{\textquotesingle}s
				toric code: A scheme for analog quantum simulation}},\ }\href
	{https://doi.org/10.1103/physrevb.104.085138} {\bibfield  {journal} {\bibinfo
			{journal} {Physical Review B}\ }\textbf {\bibinfo {volume} {104}},\ \bibinfo
		{pages} {085138} (\bibinfo {year} {2021})}\BibitemShut {NoStop}%
	\bibitem [{\citenamefont {Liu}\ \emph {et~al.}(2022)\citenamefont {Liu},
		\citenamefont {Shtengel}, \citenamefont {Smith},\ and\ \citenamefont
		{Pollmann}}]{Liu2022a}%
	\BibitemOpen
	\bibfield  {author} {\bibinfo {author} {\bibfnamefont {Y.-J.}\ \bibnamefont
			{Liu}}, \bibinfo {author} {\bibfnamefont {K.}~\bibnamefont {Shtengel}},
		\bibinfo {author} {\bibfnamefont {A.}~\bibnamefont {Smith}},\ and\ \bibinfo
		{author} {\bibfnamefont {F.}~\bibnamefont {Pollmann}},\ }\bibfield  {title}
	{\bibinfo {title} {{Methods for Simulating String-Net States and Anyons on a
				Digital Quantum Computer}},\ }\href
	{https://doi.org/10.1103/prxquantum.3.040315} {\bibfield  {journal} {\bibinfo
			{journal} {{PRX} Quantum}\ }\textbf {\bibinfo {volume} {3}},\ \bibinfo
		{pages} {040315} (\bibinfo {year} {2022})}\BibitemShut {NoStop}%
	\bibitem [{\citenamefont {Wang}\ \emph {et~al.}(2023)\citenamefont {Wang},
		\citenamefont {Aidelsburger}, \citenamefont {Dalibard}, \citenamefont
		{Eckardt},\ and\ \citenamefont {Goldman}}]{Wang2023}%
	\BibitemOpen
	\bibfield  {author} {\bibinfo {author} {\bibfnamefont {B.}~\bibnamefont
			{Wang}}, \bibinfo {author} {\bibfnamefont {M.}~\bibnamefont {Aidelsburger}},
		\bibinfo {author} {\bibfnamefont {J.}~\bibnamefont {Dalibard}}, \bibinfo
		{author} {\bibfnamefont {A.}~\bibnamefont {Eckardt}},\ and\ \bibinfo {author}
		{\bibfnamefont {N.}~\bibnamefont {Goldman}},\ }\href@noop {} {\bibinfo
		{title} {{The cold-atom elevator: From edge-state injection to the
				preparation of fractional Chern insulators}}} (\bibinfo {year} {2023}),\
	\Eprint {https://arxiv.org/abs/2306.15610} {arXiv:2306.15610} \BibitemShut
	{NoStop}%
	\bibitem [{\citenamefont {Kapit}\ \emph {et~al.}(2014)\citenamefont {Kapit},
		\citenamefont {Hafezi},\ and\ \citenamefont {Simon}}]{Kapit2014}%
	\BibitemOpen
	\bibfield  {author} {\bibinfo {author} {\bibfnamefont {E.}~\bibnamefont
			{Kapit}}, \bibinfo {author} {\bibfnamefont {M.}~\bibnamefont {Hafezi}},\ and\
		\bibinfo {author} {\bibfnamefont {S.~H.}\ \bibnamefont {Simon}},\ }\bibfield
	{title} {\bibinfo {title} {{Induced Self-Stabilization in Fractional Quantum
				Hall States of Light}},\ }\href {https://doi.org/10.1103/physrevx.4.031039}
	{\bibfield  {journal} {\bibinfo  {journal} {Physical Review X}\ }\textbf
		{\bibinfo {volume} {4}},\ \bibinfo {pages} {031039} (\bibinfo {year}
		{2014})}\BibitemShut {NoStop}%
	\bibitem [{\citenamefont {Liu}\ \emph {et~al.}(2021)\citenamefont {Liu},
		\citenamefont {Bergholtz},\ and\ \citenamefont {Budich}}]{Liu2021a}%
	\BibitemOpen
	\bibfield  {author} {\bibinfo {author} {\bibfnamefont {Z.}~\bibnamefont
			{Liu}}, \bibinfo {author} {\bibfnamefont {E.~J.}\ \bibnamefont {Bergholtz}},\
		and\ \bibinfo {author} {\bibfnamefont {J.~C.}\ \bibnamefont {Budich}},\
	}\bibfield  {title} {\bibinfo {title} {{Dissipative preparation of fractional
				Chern insulators}},\ }\href
	{https://doi.org/10.1103/PhysRevResearch.3.043119} {\bibfield  {journal}
		{\bibinfo  {journal} {Physical Review Research}\ }\textbf {\bibinfo {volume}
			{3}},\ \bibinfo {pages} {043119} (\bibinfo {year} {2021})}\BibitemShut
	{NoStop}%
	\bibitem [{\citenamefont {Kurilovich}\ \emph {et~al.}(2022)\citenamefont
		{Kurilovich}, \citenamefont {Kurilovich}, \citenamefont {Lebreuilly},\ and\
		\citenamefont {Girvin}}]{Kurilovich2022}%
	\BibitemOpen
	\bibfield  {author} {\bibinfo {author} {\bibfnamefont {P.~D.}\ \bibnamefont
			{Kurilovich}}, \bibinfo {author} {\bibfnamefont {V.~D.}\ \bibnamefont
			{Kurilovich}}, \bibinfo {author} {\bibfnamefont {J.}~\bibnamefont
			{Lebreuilly}},\ and\ \bibinfo {author} {\bibfnamefont {S.~M.}\ \bibnamefont
			{Girvin}},\ }\bibfield  {title} {\bibinfo {title} {{Stabilizing the Laughlin
				state of light: Dynamics of hole fractionalization}},\ }\bibfield  {journal}
	{\bibinfo  {journal} {{SciPost} Physics}\ }\textbf {\bibinfo {volume} {13}},\
	\href {https://doi.org/10.21468/SciPostPhys.13.5.107}
	{10.21468/SciPostPhys.13.5.107} (\bibinfo {year} {2022})\BibitemShut
	{NoStop}%
	\bibitem [{\citenamefont {Hubig}\ \emph {et~al.}(2023)\citenamefont {Hubig},
		\citenamefont {Lachenmaier}, \citenamefont {Linden}, \citenamefont
		{Reinhard}, \citenamefont {Stenzel}, \citenamefont {Swoboda}, \citenamefont
		{Grundner},\ and\ \citenamefont {Mardazad}}]{HubigSyTen}%
	\BibitemOpen
	\bibfield  {author} {\bibinfo {author} {\bibfnamefont {C.}~\bibnamefont
			{Hubig}}, \bibinfo {author} {\bibfnamefont {F.}~\bibnamefont {Lachenmaier}},
		\bibinfo {author} {\bibfnamefont {N.-O.}\ \bibnamefont {Linden}}, \bibinfo
		{author} {\bibfnamefont {T.}~\bibnamefont {Reinhard}}, \bibinfo {author}
		{\bibfnamefont {L.}~\bibnamefont {Stenzel}}, \bibinfo {author} {\bibfnamefont
			{A.}~\bibnamefont {Swoboda}}, \bibinfo {author} {\bibfnamefont
			{M.}~\bibnamefont {Grundner}},\ and\ \bibinfo {author} {\bibfnamefont
			{S.}~\bibnamefont {Mardazad}},\ }\href {https://syten.eu} {\bibinfo {title}
		{The \textsc{SyTen} toolkit}} (\bibinfo {year} {2023})\BibitemShut {NoStop}%
	\bibitem [{\citenamefont {White}(1992)}]{White1992}%
	\BibitemOpen
	\bibfield  {author} {\bibinfo {author} {\bibfnamefont {S.~R.}\ \bibnamefont
			{White}},\ }\bibfield  {title} {\bibinfo {title} {{Density matrix formulation
				for quantum renormalization groups}},\ }\href
	{https://doi.org/10.1103/physrevlett.69.2863} {\bibfield  {journal} {\bibinfo
			{journal} {Physical Review Letters}\ }\textbf {\bibinfo {volume} {69}},\
		\bibinfo {pages} {2863} (\bibinfo {year} {1992})}\BibitemShut {NoStop}%
	\bibitem [{\citenamefont {Schollwöck}(2005)}]{Schollwoeck2005}%
	\BibitemOpen
	\bibfield  {author} {\bibinfo {author} {\bibfnamefont {U.}~\bibnamefont
			{Schollwöck}},\ }\bibfield  {title} {\bibinfo {title} {The density-matrix
			renormalization group},\ }\href {https://doi.org/10.1103/revmodphys.77.259}
	{\bibfield  {journal} {\bibinfo  {journal} {Reviews of Modern Physics}\
		}\textbf {\bibinfo {volume} {77}},\ \bibinfo {pages} {259} (\bibinfo {year}
		{2005})}\BibitemShut {NoStop}%
	\bibitem [{\citenamefont {Schollwöck}(2011)}]{Schollwoeck2011}%
	\BibitemOpen
	\bibfield  {author} {\bibinfo {author} {\bibfnamefont {U.}~\bibnamefont
			{Schollwöck}},\ }\bibfield  {title} {\bibinfo {title} {{The density-matrix
				renormalization group in the age of matrix product states}},\ }\href
	{https://doi.org/10.1016/j.aop.2010.09.012} {\bibfield  {journal} {\bibinfo
			{journal} {Annals of Physics}\ }\textbf {\bibinfo {volume} {326}},\ \bibinfo
		{pages} {96} (\bibinfo {year} {2011})}\BibitemShut {NoStop}%
	\bibitem [{\citenamefont {Haegeman}\ \emph {et~al.}(2011)\citenamefont
		{Haegeman}, \citenamefont {Cirac}, \citenamefont {Osborne}, \citenamefont
		{Pi{\v{z}}orn}, \citenamefont {Verschelde},\ and\ \citenamefont
		{Verstraete}}]{Haegeman2011}%
	\BibitemOpen
	\bibfield  {author} {\bibinfo {author} {\bibfnamefont {J.}~\bibnamefont
			{Haegeman}}, \bibinfo {author} {\bibfnamefont {J.~I.}\ \bibnamefont {Cirac}},
		\bibinfo {author} {\bibfnamefont {T.~J.}\ \bibnamefont {Osborne}}, \bibinfo
		{author} {\bibfnamefont {I.}~\bibnamefont {Pi{\v{z}}orn}}, \bibinfo {author}
		{\bibfnamefont {H.}~\bibnamefont {Verschelde}},\ and\ \bibinfo {author}
		{\bibfnamefont {F.}~\bibnamefont {Verstraete}},\ }\bibfield  {title}
	{\bibinfo {title} {{Time-Dependent Variational Principle for Quantum
				Lattices}},\ }\href {https://doi.org/10.1103/physrevlett.107.070601}
	{\bibfield  {journal} {\bibinfo  {journal} {Physical Review Letters}\
		}\textbf {\bibinfo {volume} {107}},\ \bibinfo {pages} {070601} (\bibinfo
		{year} {2011})}\BibitemShut {NoStop}%
	\bibitem [{\citenamefont {Haegeman}\ \emph {et~al.}(2016)\citenamefont
		{Haegeman}, \citenamefont {Lubich}, \citenamefont {Oseledets}, \citenamefont
		{Vandereycken},\ and\ \citenamefont {Verstraete}}]{Haegeman2016}%
	\BibitemOpen
	\bibfield  {author} {\bibinfo {author} {\bibfnamefont {J.}~\bibnamefont
			{Haegeman}}, \bibinfo {author} {\bibfnamefont {C.}~\bibnamefont {Lubich}},
		\bibinfo {author} {\bibfnamefont {I.}~\bibnamefont {Oseledets}}, \bibinfo
		{author} {\bibfnamefont {B.}~\bibnamefont {Vandereycken}},\ and\ \bibinfo
		{author} {\bibfnamefont {F.}~\bibnamefont {Verstraete}},\ }\bibfield  {title}
	{\bibinfo {title} {Unifying time evolution and optimization with matrix
			product states},\ }\href {https://doi.org/10.1103/physrevb.94.165116}
	{\bibfield  {journal} {\bibinfo  {journal} {Physical Review B}\ }\textbf
		{\bibinfo {volume} {94}},\ \bibinfo {pages} {165116} (\bibinfo {year}
		{2016})}\BibitemShut {NoStop}%
	\bibitem [{\citenamefont {St\v{r}eda}(1982{\natexlab{a}})}]{Streda1982}%
	\BibitemOpen
	\bibfield  {author} {\bibinfo {author} {\bibfnamefont {P.}~\bibnamefont
			{St\v{r}eda}},\ }\bibfield  {title} {\bibinfo {title} {{Theory of quantised
				Hall conductivity in two dimensions}},\ }\href
	{https://doi.org/10.1088/0022-3719/15/22/005} {\bibfield  {journal} {\bibinfo
			{journal} {Journal of Physics C: Solid State Physics}\ }\textbf {\bibinfo
			{volume} {15}},\ \bibinfo {pages} {L717} (\bibinfo {year}
		{1982}{\natexlab{a}})}\BibitemShut {NoStop}%
	\bibitem [{\citenamefont {St\v{r}eda}(1982{\natexlab{b}})}]{Streda1982a}%
	\BibitemOpen
	\bibfield  {author} {\bibinfo {author} {\bibfnamefont {P.}~\bibnamefont
			{St\v{r}eda}},\ }\bibfield  {title} {\bibinfo {title} {{Quantised Hall effect
				in a two-dimensional periodic potential}},\ }\href
	{https://doi.org/10.1088/0022-3719/15/36/006} {\bibfield  {journal} {\bibinfo
			{journal} {Journal of Physics C: Solid State Physics}\ }\textbf {\bibinfo
			{volume} {15}},\ \bibinfo {pages} {L1299} (\bibinfo {year}
		{1982}{\natexlab{b}})}\BibitemShut {NoStop}%
	\bibitem [{\citenamefont {Repellin}\ \emph {et~al.}(2020)\citenamefont
		{Repellin}, \citenamefont {L{\'{e}}onard},\ and\ \citenamefont
		{Goldman}}]{Repellin2020}%
	\BibitemOpen
	\bibfield  {author} {\bibinfo {author} {\bibfnamefont {C.}~\bibnamefont
			{Repellin}}, \bibinfo {author} {\bibfnamefont {J.}~\bibnamefont
			{L{\'{e}}onard}},\ and\ \bibinfo {author} {\bibfnamefont {N.}~\bibnamefont
			{Goldman}},\ }\bibfield  {title} {\bibinfo {title} {{Fractional Chern
				insulators of few bosons in a box: Hall plateaus from center-of-mass drifts
				and density profiles}},\ }\href {https://doi.org/10.1103/physreva.102.063316}
	{\bibfield  {journal} {\bibinfo  {journal} {Physical Review A}\ }\textbf
		{\bibinfo {volume} {102}},\ \bibinfo {pages} {063316} (\bibinfo {year}
		{2020})}\BibitemShut {NoStop}%
	\bibitem [{\citenamefont {Zupancic}\ \emph {et~al.}(2016)\citenamefont
		{Zupancic}, \citenamefont {Preiss}, \citenamefont {Ma}, \citenamefont
		{Lukin}, \citenamefont {Tai}, \citenamefont {Rispoli}, \citenamefont
		{Islam},\ and\ \citenamefont {Greiner}}]{Zupancic2016}%
	\BibitemOpen
	\bibfield  {author} {\bibinfo {author} {\bibfnamefont {P.}~\bibnamefont
			{Zupancic}}, \bibinfo {author} {\bibfnamefont {P.~M.}\ \bibnamefont
			{Preiss}}, \bibinfo {author} {\bibfnamefont {R.}~\bibnamefont {Ma}}, \bibinfo
		{author} {\bibfnamefont {A.}~\bibnamefont {Lukin}}, \bibinfo {author}
		{\bibfnamefont {M.~E.}\ \bibnamefont {Tai}}, \bibinfo {author} {\bibfnamefont
			{M.}~\bibnamefont {Rispoli}}, \bibinfo {author} {\bibfnamefont
			{R.}~\bibnamefont {Islam}},\ and\ \bibinfo {author} {\bibfnamefont
			{M.}~\bibnamefont {Greiner}},\ }\bibfield  {title} {\bibinfo {title}
		{{Ultra-precise holographic beam shaping for microscopic quantum control}},\
	}\href {https://doi.org/10.1364/oe.24.013881} {\bibfield  {journal} {\bibinfo
			{journal} {Optics Express}\ }\textbf {\bibinfo {volume} {24}},\ \bibinfo
		{pages} {13881} (\bibinfo {year} {2016})}\BibitemShut {NoStop}%
	\bibitem [{\citenamefont {Gaunt}\ and\ \citenamefont
		{Hadzibabic}(2012)}]{Gaunt2012}%
	\BibitemOpen
	\bibfield  {author} {\bibinfo {author} {\bibfnamefont {A.~L.}\ \bibnamefont
			{Gaunt}}\ and\ \bibinfo {author} {\bibfnamefont {Z.}~\bibnamefont
			{Hadzibabic}},\ }\bibfield  {title} {\bibinfo {title} {{Robust Digital
				Holography For Ultracold Atom Trapping}},\ }\bibfield  {journal} {\bibinfo
		{journal} {Scientific Reports}\ }\textbf {\bibinfo {volume} {2}},\ \href
	{https://doi.org/10.1038/srep00721} {10.1038/srep00721} (\bibinfo {year}
	{2012})\BibitemShut {NoStop}%
	\bibitem [{\citenamefont {Eisert}\ \emph {et~al.}(2010)\citenamefont {Eisert},
		\citenamefont {Cramer},\ and\ \citenamefont {Plenio}}]{Eisert2010}%
	\BibitemOpen
	\bibfield  {author} {\bibinfo {author} {\bibfnamefont {J.}~\bibnamefont
			{Eisert}}, \bibinfo {author} {\bibfnamefont {M.}~\bibnamefont {Cramer}},\
		and\ \bibinfo {author} {\bibfnamefont {M.~B.}\ \bibnamefont {Plenio}},\
	}\bibfield  {title} {\bibinfo {title} {{Colloquium: Area laws for the
				entanglement entropy}},\ }\href {https://doi.org/10.1103/revmodphys.82.277}
	{\bibfield  {journal} {\bibinfo  {journal} {Reviews of Modern Physics}\
		}\textbf {\bibinfo {volume} {82}},\ \bibinfo {pages} {277} (\bibinfo {year}
		{2010})}\BibitemShut {NoStop}%
	\bibitem [{\citenamefont {Cirac}\ \emph {et~al.}(2021)\citenamefont {Cirac},
		\citenamefont {P\'{e}rez-Garc'{i}a}, \citenamefont {Schuch},\ and\
		\citenamefont {Verstraete}}]{Cirac2021}%
	\BibitemOpen
	\bibfield  {author} {\bibinfo {author} {\bibfnamefont {J.~I.}\ \bibnamefont
			{Cirac}}, \bibinfo {author} {\bibfnamefont {D.}~\bibnamefont
			{P\'{e}rez-Garc'{i}a}}, \bibinfo {author} {\bibfnamefont {N.}~\bibnamefont
			{Schuch}},\ and\ \bibinfo {author} {\bibfnamefont {F.}~\bibnamefont
			{Verstraete}},\ }\bibfield  {title} {\bibinfo {title} {Matrix product states
			and projected entangled pair states: Concepts, symmetries, theorems},\ }\href
	{https://doi.org/10.1103/revmodphys.93.045003} {\bibfield  {journal}
		{\bibinfo  {journal} {Reviews of Modern Physics}\ }\textbf {\bibinfo {volume}
			{93}},\ \bibinfo {pages} {045003} (\bibinfo {year} {2021})}\BibitemShut
	{NoStop}%
	\bibitem [{\citenamefont {Ra{\v{c}}i{\={u}}nas}\ \emph
		{et~al.}(2018)\citenamefont {Ra{\v{c}}i{\={u}}nas}, \citenamefont {\"Unal},
		\citenamefont {Anisimovas},\ and\ \citenamefont {Eckardt}}]{Raciunas2018}%
	\BibitemOpen
	\bibfield  {author} {\bibinfo {author} {\bibfnamefont {M.}~\bibnamefont
			{Ra{\v{c}}i{\={u}}nas}}, \bibinfo {author} {\bibfnamefont {F.~N.}\
			\bibnamefont {\"Unal}}, \bibinfo {author} {\bibfnamefont {E.}~\bibnamefont
			{Anisimovas}},\ and\ \bibinfo {author} {\bibfnamefont {A.}~\bibnamefont
			{Eckardt}},\ }\bibfield  {title} {\bibinfo {title} {{Creating, probing, and
				manipulating fractionally charged excitations of fractional Chern insulators
				in optical lattices}},\ }\href {https://doi.org/10.1103/physreva.98.063621}
	{\bibfield  {journal} {\bibinfo  {journal} {Physical Review A}\ }\textbf
		{\bibinfo {volume} {98}},\ \bibinfo {pages} {063621} (\bibinfo {year}
		{2018})}\BibitemShut {NoStop}%
	\bibitem [{\citenamefont {Wang}\ \emph {et~al.}(2022)\citenamefont {Wang},
		\citenamefont {Dong},\ and\ \citenamefont {Eckardt}}]{Wang2022}%
	\BibitemOpen
	\bibfield  {author} {\bibinfo {author} {\bibfnamefont {B.}~\bibnamefont
			{Wang}}, \bibinfo {author} {\bibfnamefont {X.}~\bibnamefont {Dong}},\ and\
		\bibinfo {author} {\bibfnamefont {A.}~\bibnamefont {Eckardt}},\ }\bibfield
	{title} {\bibinfo {title} {{Measurable signatures of bosonic fractional Chern
				insulator states and their fractional excitations in a quantum-gas
				microscope}},\ }\href {https://doi.org/10.21468/scipostphys.12.3.095}
	{\bibfield  {journal} {\bibinfo  {journal} {{SciPost} Physics}\ }\textbf
		{\bibinfo {volume} {12}},\ \bibinfo {pages} {095} (\bibinfo {year}
		{2022})}\BibitemShut {NoStop}%
	\bibitem [{\citenamefont {Kivelson}\ and\ \citenamefont
		{Schrieffer}(1982)}]{Kivelson1982}%
	\BibitemOpen
	\bibfield  {author} {\bibinfo {author} {\bibfnamefont {S.}~\bibnamefont
			{Kivelson}}\ and\ \bibinfo {author} {\bibfnamefont {J.~R.}\ \bibnamefont
			{Schrieffer}},\ }\bibfield  {title} {\bibinfo {title} {Fractional charge, a
			sharp quantum observable},\ }\href {https://doi.org/10.1103/physrevb.25.6447}
	{\bibfield  {journal} {\bibinfo  {journal} {Physical Review B}\ }\textbf
		{\bibinfo {volume} {25}},\ \bibinfo {pages} {6447} (\bibinfo {year}
		{1982})}\BibitemShut {NoStop}%
	\bibitem [{\citenamefont {Bibo}\ \emph {et~al.}(2020)\citenamefont {Bibo},
		\citenamefont {Lovas}, \citenamefont {You}, \citenamefont {Grusdt},\ and\
		\citenamefont {Pollmann}}]{Bibo2020}%
	\BibitemOpen
	\bibfield  {author} {\bibinfo {author} {\bibfnamefont {J.}~\bibnamefont
			{Bibo}}, \bibinfo {author} {\bibfnamefont {I.}~\bibnamefont {Lovas}},
		\bibinfo {author} {\bibfnamefont {Y.}~\bibnamefont {You}}, \bibinfo {author}
		{\bibfnamefont {F.}~\bibnamefont {Grusdt}},\ and\ \bibinfo {author}
		{\bibfnamefont {F.}~\bibnamefont {Pollmann}},\ }\bibfield  {title} {\bibinfo
		{title} {{Fractional corner charges in a two-dimensional superlattice
				Bose-Hubbard model}},\ }\href {https://doi.org/10.1103/physrevb.102.041126}
	{\bibfield  {journal} {\bibinfo  {journal} {Physical Review B}\ }\textbf
		{\bibinfo {volume} {102}},\ \bibinfo {pages} {041126(R)} (\bibinfo {year}
		{2020})}\BibitemShut {NoStop}%
	\bibitem [{\citenamefont {Ferris}\ and\ \citenamefont
		{Vidal}(2012)}]{Ferris2012}%
	\BibitemOpen
	\bibfield  {author} {\bibinfo {author} {\bibfnamefont {A.~J.}\ \bibnamefont
			{Ferris}}\ and\ \bibinfo {author} {\bibfnamefont {G.}~\bibnamefont {Vidal}},\
	}\bibfield  {title} {\bibinfo {title} {Perfect sampling with unitary tensor
			networks},\ }\href {https://doi.org/10.1103/physrevb.85.165146} {\bibfield
		{journal} {\bibinfo  {journal} {Physical Review B}\ }\textbf {\bibinfo
			{volume} {85}},\ \bibinfo {pages} {165146} (\bibinfo {year}
		{2012})}\BibitemShut {NoStop}%
	\bibitem [{\citenamefont {Wigley}\ \emph {et~al.}(2016)\citenamefont {Wigley},
		\citenamefont {Everitt}, \citenamefont {van~den Hengel}, \citenamefont
		{Bastian}, \citenamefont {Sooriyabandara}, \citenamefont {McDonald},
		\citenamefont {Hardman}, \citenamefont {Quinlivan}, \citenamefont {Manju},
		\citenamefont {Kuhn}, \citenamefont {Petersen}, \citenamefont {Luiten},
		\citenamefont {Hope}, \citenamefont {Robins},\ and\ \citenamefont
		{Hush}}]{Wigley2016}%
	\BibitemOpen
	\bibfield  {author} {\bibinfo {author} {\bibfnamefont {P.~B.}\ \bibnamefont
			{Wigley}}, \bibinfo {author} {\bibfnamefont {P.~J.}\ \bibnamefont {Everitt}},
		\bibinfo {author} {\bibfnamefont {A.}~\bibnamefont {van~den Hengel}},
		\bibinfo {author} {\bibfnamefont {J.~W.}\ \bibnamefont {Bastian}}, \bibinfo
		{author} {\bibfnamefont {M.~A.}\ \bibnamefont {Sooriyabandara}}, \bibinfo
		{author} {\bibfnamefont {G.~D.}\ \bibnamefont {McDonald}}, \bibinfo {author}
		{\bibfnamefont {K.~S.}\ \bibnamefont {Hardman}}, \bibinfo {author}
		{\bibfnamefont {C.~D.}\ \bibnamefont {Quinlivan}}, \bibinfo {author}
		{\bibfnamefont {P.}~\bibnamefont {Manju}}, \bibinfo {author} {\bibfnamefont
			{C.~C.~N.}\ \bibnamefont {Kuhn}}, \bibinfo {author} {\bibfnamefont {I.~R.}\
			\bibnamefont {Petersen}}, \bibinfo {author} {\bibfnamefont {A.~N.}\
			\bibnamefont {Luiten}}, \bibinfo {author} {\bibfnamefont {J.~J.}\
			\bibnamefont {Hope}}, \bibinfo {author} {\bibfnamefont {N.~P.}\ \bibnamefont
			{Robins}},\ and\ \bibinfo {author} {\bibfnamefont {M.~R.}\ \bibnamefont
			{Hush}},\ }\bibfield  {title} {\bibinfo {title} {Fast machine-learning online
			optimization of ultra-cold-atom experiments},\ }\href
	{https://doi.org/10.1038/srep25890} {\bibfield  {journal} {\bibinfo
			{journal} {Scientific Reports}\ }\textbf {\bibinfo {volume} {6}},\ \bibinfo
		{pages} {28632} (\bibinfo {year} {2016})}\BibitemShut {NoStop}%
	\bibitem [{\citenamefont {Vendeiro}\ \emph {et~al.}(2022)\citenamefont
		{Vendeiro}, \citenamefont {Ramette}, \citenamefont {Rudelis}, \citenamefont
		{Chong}, \citenamefont {Sinclair}, \citenamefont {Stewart}, \citenamefont
		{Urvoy},\ and\ \citenamefont {Vuleti{\'{c}}}}]{Vendeiro2022}%
	\BibitemOpen
	\bibfield  {author} {\bibinfo {author} {\bibfnamefont {Z.}~\bibnamefont
			{Vendeiro}}, \bibinfo {author} {\bibfnamefont {J.}~\bibnamefont {Ramette}},
		\bibinfo {author} {\bibfnamefont {A.}~\bibnamefont {Rudelis}}, \bibinfo
		{author} {\bibfnamefont {M.}~\bibnamefont {Chong}}, \bibinfo {author}
		{\bibfnamefont {J.}~\bibnamefont {Sinclair}}, \bibinfo {author}
		{\bibfnamefont {L.}~\bibnamefont {Stewart}}, \bibinfo {author} {\bibfnamefont
			{A.}~\bibnamefont {Urvoy}},\ and\ \bibinfo {author} {\bibfnamefont
			{V.}~\bibnamefont {Vuleti{\'{c}}}},\ }\bibfield  {title} {\bibinfo {title}
		{{Machine-learning-accelerated Bose-Einstein condensation}},\ }\href
	{https://doi.org/10.1103/physrevresearch.4.043216} {\bibfield  {journal}
		{\bibinfo  {journal} {Physical Review Research}\ }\textbf {\bibinfo {volume}
			{4}},\ \bibinfo {pages} {043216} (\bibinfo {year} {2022})}\BibitemShut
	{NoStop}%
	\bibitem [{\citenamefont {Xie}\ \emph {et~al.}(2022)\citenamefont {Xie},
		\citenamefont {Dai}, \citenamefont {Yuan}, \citenamefont {Deng},
		\citenamefont {Li}, \citenamefont {Chen},\ and\ \citenamefont
		{Pan}}]{Xie2022}%
	\BibitemOpen
	\bibfield  {author} {\bibinfo {author} {\bibfnamefont {Y.-J.}\ \bibnamefont
			{Xie}}, \bibinfo {author} {\bibfnamefont {H.-N.}\ \bibnamefont {Dai}},
		\bibinfo {author} {\bibfnamefont {Z.-S.}\ \bibnamefont {Yuan}}, \bibinfo
		{author} {\bibfnamefont {Y.}~\bibnamefont {Deng}}, \bibinfo {author}
		{\bibfnamefont {X.}~\bibnamefont {Li}}, \bibinfo {author} {\bibfnamefont
			{Y.-A.}\ \bibnamefont {Chen}},\ and\ \bibinfo {author} {\bibfnamefont
			{J.-W.}\ \bibnamefont {Pan}},\ }\bibfield  {title} {\bibinfo {title}
		{Bayesian learning for optimal control of quantum many-body states in optical
			lattices},\ }\href {https://doi.org/10.1103/physreva.106.013316} {\bibfield
		{journal} {\bibinfo  {journal} {Physical Review A}\ }\textbf {\bibinfo
			{volume} {106}},\ \bibinfo {pages} {013316} (\bibinfo {year}
		{2022})}\BibitemShut {NoStop}%
	\bibitem [{\citenamefont {Blatz}\ \emph {et~al.}(2023)\citenamefont {Blatz},
		\citenamefont {Kwan}, \citenamefont {Léonard},\ and\ \citenamefont
		{Bohrdt}}]{Blatz2023}%
	\BibitemOpen
	\bibfield  {author} {\bibinfo {author} {\bibfnamefont {T.}~\bibnamefont
			{Blatz}}, \bibinfo {author} {\bibfnamefont {J.}~\bibnamefont {Kwan}},
		\bibinfo {author} {\bibfnamefont {J.}~\bibnamefont {Léonard}},\ and\
		\bibinfo {author} {\bibfnamefont {A.}~\bibnamefont {Bohrdt}},\ }\bibfield
	{title} {\bibinfo {title} {Bayesian optimization for robust state preparation
			in quantum many-body systems},\ }\href@noop {} {\  (\bibinfo {year}
		{2023})},\ \Eprint {https://arxiv.org/abs/2312.09253} {arXiv:2312.09253}
	\BibitemShut {NoStop}%
\end{thebibliography}

%

\end{document}